\documentclass[10pt,conference]{IEEEtran}

\usepackage{amsmath}
\usepackage{amssymb}
\usepackage{booktabs}
\usepackage{enumitem}
\usepackage{hyperref}
\usepackage{xcolor}
\usepackage{tikz}
\usepackage{graphicx}
\usepackage{eso-pic}
\usetikzlibrary{arrows.meta,positioning,shapes.geometric,fit,calc}

\hypersetup{
  colorlinks=true,
  linkcolor=blue,
  citecolor=blue,
  urlcolor=blue
}

\title{OpenAgenet / OAN Yellow Paper: Technical Architecture for Trust-Governed Resource Identity and Discovery}

\author{
\IEEEauthorblockN{Jinliang Xu}
\IEEEauthorblockA{\textit{China Academy of Information and Communications Technology}, Beijing, China \\
xujinliang@caict.ac.cn; jlxufly@gmail.com}
}

\begin{document}

\AddToShipoutPictureBG*{%
  \begin{tikzpicture}[remember picture,overlay]
    \node[opacity=0.22] at (current page.center) {%
      \includegraphics[width=.333\paperwidth,keepaspectratio]{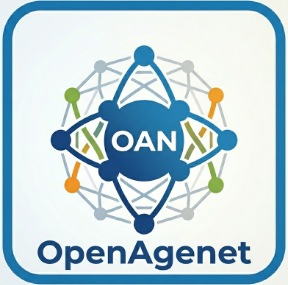}%
    };
  \end{tikzpicture}%
}

\maketitle

\begin{abstract}
This yellow paper specifies the technical architecture and verification
semantics of OpenAgenet (OAN), a protocol-neutral trust layer for heterogeneous
Agent resources. The design combines \texttt{did:oan} decentralized identifiers
(DIDs), verifiable credentials (VCs), on-chain infrastructure governance, and
Root-verified resource packages. It defines the objects and lifecycle
transitions that connect subject-control verification, Registrar admission,
Root acceptance, authorized Discovery, and signed invocation. Infrastructure
authorization requires both active governance state and a valid Root-issued
credential; resource discovery additionally requires package integrity,
current-version state, and authorized-domain coverage. Local OAN Identity
signing material remains separate from public protocol objects. Integration
profiles carry resource identity into the Model Context Protocol (MCP) and
Agent-to-Agent (A2A) protocol through Skill, Runtime, Sidecar, and Gateway
interfaces without redefining native task semantics. The paper distinguishes
protocol requirements from implementation choices and sets out security
properties, persistence and recovery expectations, and conformance criteria.
Together, these rules provide a technical basis for independently implemented
registration, discovery, and identity-protected interaction within an OAN trust
domain.
\end{abstract}

\noindent\textbf{Public portal:}
\href{https://www.openagenet.xyz/}{https://www.openagenet.xyz/}

\noindent\textbf{Open-source project:}
\href{https://github.com/OpenAgenet}{https://github.com/OpenAgenet}


\section{Technical Scope}

\begin{figure}[!ht]
\centering
\includegraphics[width=\columnwidth]{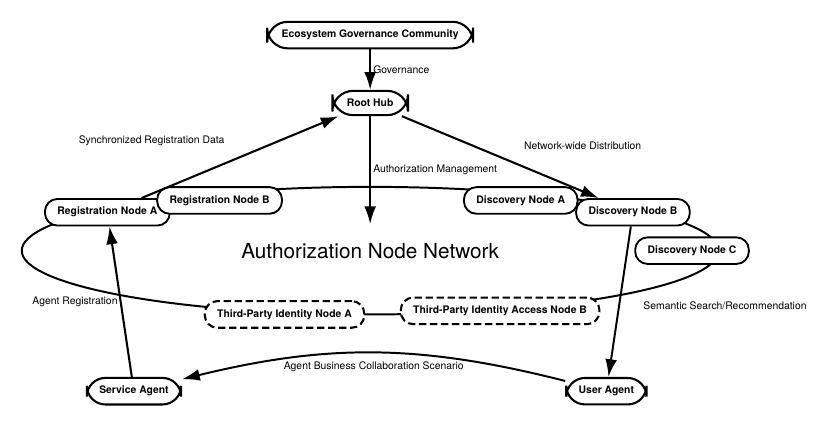}
\caption{OAN technical role framework. Governance authorizes infrastructure
roles, Root enforces trust and publication authority, Registrars submit
resources, Discovery nodes return verified candidates, and Agents invoke after
verification.}
\label{fig:yp-oan-framework}
\end{figure}

OpenAgenet (OAN) uses decentralized identifiers (DIDs) and verifiable
credentials (VCs) within a trust-governed resource identity and Discovery
architecture. It specifies how \texttt{did:oan} identity representations become
admissible across registration, Root acceptance, package distribution,
Discovery indexing, response verification, and pre-connection invocation or
access. Task protocols, model orchestration, prompt strategy, tool execution,
ranking, artifact hosting, and business ontologies remain application-layer
concerns.
Figure~\ref{fig:yp-oan-framework} provides the role-level orientation for the
technical profiles and verification rules defined below.

The technical scope can be expressed through the following invariant: a
resource identity should become visible to a relying party only after it has
been prepared by the resource provider, checked by a Registrar whose
infrastructure status is active under governance and credentialed by Root,
accepted by Root, distributed through a verifiable package, indexed by a
Discovery node whose infrastructure status is also active and credentialed,
and returned with evidence that can be checked before invocation or access.
Root anchors accepted trust evidence, publishes verified resource state, and
governs tag-tree vocabulary and version compatibility for registration and
Discovery. Semantic query execution remains a Discovery-node function.

The core abstraction is a \textbf{Trusted Resource}, which includes Agents and
other AI resource types. A trusted resource is represented by a \texttt{did:oan} DID
Document and a Root-verified ResourcePackage, whose lifecycle and discovery
evidence can be checked before use. The initial resource types are Agent
Service, Skill, Model Context Protocol (MCP) Server, and Tool/application
programming interface (API). This
abstraction lets OAN implement a
Resource First model: the protocol admits heterogeneous AI ecosystem assets
through one trust path, while allowing each product type to keep its native
runtime, artifact format, or interaction protocol.

Verification requirements, authorization rules, lifecycle transitions, and
failure behavior define OAN's protocol contracts. Deployment topology, storage
backends, queueing, indexing, and public interfaces are implementation choices.
Conforming implementations preserve verifiable resource identity, governed
admission, Root-verified publication, scoped Discovery, and pre-connection
verification across these choices.

\section{OAN Protocol Stack}

OAN comprises a set of protocol profiles with distinct responsibilities and
dependencies. Operators can implement and combine the profiles required by
their roles. The lower layers reuse the DID/VC family of
identity and credential concepts~\cite{w3c_did_core,vc_data_model_2,vc_data_integrity,did_resolution},
while its upper layers are designed to wrap Agent interaction protocols such as
MCP, Agent-to-Agent (A2A), and Agent Network Protocol
(ANP)~\cite{mcp_spec,a2a_spec,did_wba_spec}. Package distribution uses
a content delivery network (CDN), and software development kits (SDKs) expose
these profiles to application developers. Table~\ref{tab:oan-stack} summarizes
the objects and users associated with each profile.

\begin{table}[t]
\centering
\caption{OAN Protocol Stack}
\label{tab:oan-stack}
\scriptsize
\begin{tabular}{@{}p{0.18\columnwidth}p{0.42\columnwidth}p{0.28\columnwidth}@{}}
\toprule
Profile & Main objects & Primary users \\
\midrule
Identity & \texttt{did:oan} DID, DID Document, OAN Identity, resource metadata & SDKs \\
Registration & draft, proof, credential & Registrar, Root \\
Federated governance & lifecycle state, threshold events & committee, Root \\
Root enforcement & Root VC, accepted record, package proof & Root, infrastructure \\
Package & verified package, cursor & Root, CDN, Discovery \\
Discovery & query, signed response & Discovery, User Agent \\
Invocation/\allowbreak access & signed request/response, artifact reference & consumers/providers \\
Adapter & MCP/A2A/ANP mapping & protocol integrators \\
\bottomrule
\end{tabular}
\end{table}

The stack structure clarifies implementation scope. A resource provider may
only need the Identity, Discovery, and Invocation/access profiles. A Discovery
operator needs Package and Discovery profiles plus Root bulletin verification.
A full infrastructure operator needs Registration, Root Enforcement, Package,
and Discovery. Governance operation clients additionally implement the
Federated Governance profile. An adjacent protocol community can define an
Adapter profile without changing the Root acceptance model.

\begin{figure*}[!ht]
\centering
\includegraphics[width=0.7\linewidth]{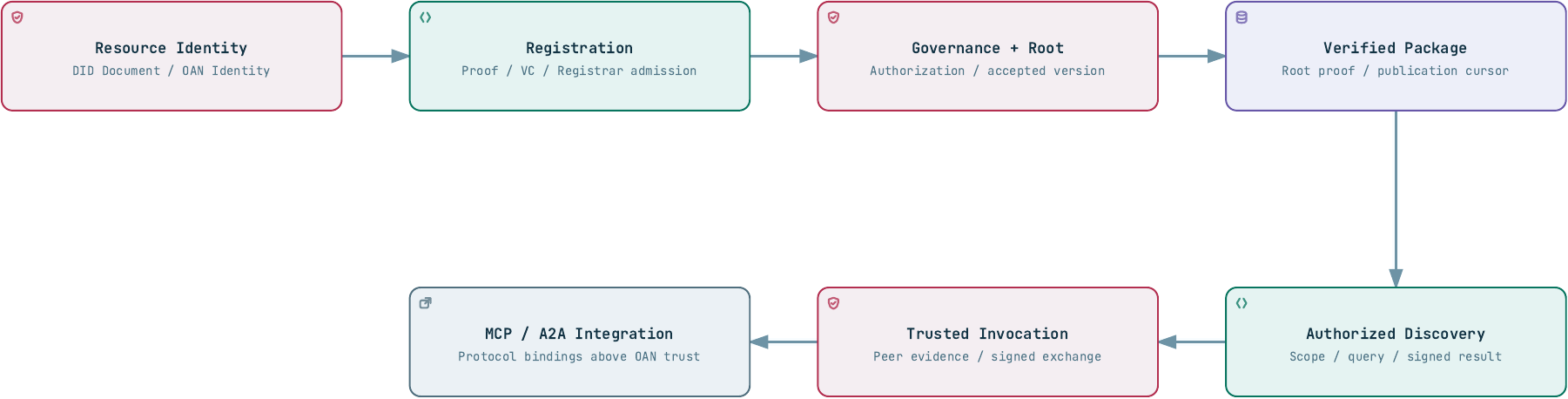}
\caption{OAN technical profiles and their dependencies, from resource identity and registration to discovery, invocation, and protocol adaptation.}
\label{fig:yellowpaper-stack}
\end{figure*}

Figure~\ref{fig:yellowpaper-stack} shows how these profiles connect resource
identity and registration to publication, discovery, invocation, and adaptation.

\section{System Assumptions}

OAN assumes one Root service boundary per trust domain, with infrastructure
governance separated from Root enforcement. Infrastructure lifecycle state for Registrar,
Discovery, and future VC issuer nodes can be governed above Root by a committee
or threshold process and published through the on-chain governance layer. Root
consumes this state, enforces it, and issues protocol credentials. This
architecture establishes a governed trust-domain profile. Interoperable
federation among independent domains and Roots requires future profiles and
conformance evidence.
Within a trust domain, relying parties ultimately trust the Root signing
boundary to represent accepted state correctly. On-chain governance constrains
infrastructure authorization, but it does not make a compromised Root key or a
captured governance process cryptographically distinguishable from legitimate
trust-domain action.

OAN also assumes that participants can perform standard cryptographic
verification. Agents and infrastructure nodes are expected to hold DID-style
identity material and signing keys. Signed objects use canonical serialization
and hash binding. Time synchronization is assumed to be good enough for
timestamp freshness windows, while nonce storage handles replay protection.
The trust posture is aligned with zero-trust and workload-identity principles:
every privileged role and runtime request must be explicitly verifiable rather
than implicitly accepted by network location~\cite{rose2020zerotrust,spiffe_spire}.

The protocol-facing trust path should not depend on private review evidence
such as email, phone, web account login, or manually reviewed forms. Such
evidence may support governance workflow, but runtime verification should rely
on DID documents, credentials, signed envelopes, Root bulletin facts,
Root-verified packages, and Discovery response proofs.

\section{Trust Layers}

OAN separates social governance, admission workflow, and runtime protocol
verification. The layers are shown in Table~\ref{tab:trust-layers}.

\begin{table}[t]
\caption{OAN Trust Layers}
\label{tab:trust-layers}
\centering
\resizebox{\columnwidth}{!}{%
\begin{tabular}{@{}lll@{}}
\toprule
Layer & Purpose & Mechanism \\
\midrule
Federated governance & infrastructure lifecycle & threshold event \\
Root enforcement & protocol authorization & Root-issued VC \\
Registration review & resource admission workflow & Registrar credential \\
Subject control & prove DID ownership & challenge signature \\
Root acceptance & admit identity version & signed Root record \\
Package distribution & publish identity state & Root proof and hash \\
Semantic governance & govern tag tree & versioned taxonomy state \\
Discovery exposure & scoped query visibility & domain authorization \\
Runtime invocation & verify request peer & signed envelope \\
\bottomrule
\end{tabular}
}
\end{table}

This separation makes failures easier to reason about. If a Registrar is
misconfigured, Root still checks Registrar authorization and credential proof.
If a CDN returns a tampered package, Discovery recomputes hashes and verifies
Root proof. If a Discovery node is stale or unauthorized, User Agents can reject
responses that fail freshness or authorization checks. If a caller replays an
old invocation, the Service Agent can reject it using nonce and timestamp
validation.

The threshold event in the first layer is a governance fact, not a protocol
credential. A service node becomes protocol-authorized only after Root verifies
the governance fact, validates the node's DID-control evidence, and issues the
corresponding infrastructure authorization VC.
The semantic-governance layer is also Root-mediated, but it is not a request
execution path. It defines the governed semantic tag tree that Registrars use
for registration-time guidance and that Discovery nodes use to maintain
compatible semantic indexes.

\section{Core Entities and Notation}

Let $\mathcal{A}$ be Agent subjects, $\mathcal{R}$ Registrar nodes,
$\mathcal{D}$ Discovery nodes, $\mathcal{V}$ third-party VC issuer nodes,
$\gamma$ the governance layer, and $\rho$ the Root service in one trust
domain. For an infrastructure subject $s \in \mathcal{R}\cup\mathcal{D}\cup
\mathcal{V}$, let $G_\gamma(s)$ be the latest governance status and
$VC_\rho(s)$ be the Root-issued infrastructure authorization VC. Let
$AD(x)$ be the typed authorized-domain set associated with subject or resource
$x$. The subject is protocol-authorized only if:

\[
  InfraAuth(s) = Active(G_\gamma(s)) \land Valid(VC_\rho(s)).
\]

This predicate is deliberately conjunctive. Governance approval without a Root
VC is not yet a protocol credential, and a stale Root VC without active
governance state is not sufficient for new infrastructure service.
For Registrar and Discovery subjects, $AD(s)$ is also part of the governance
and VC-bound authorization state. For VC issuer subjects it remains empty
unless a later profile gives that role a domain-scoped authority.

A resource $r$ has a versioned identity representation $M_r^v$. The
representation hash is $H(M_r^v)$ and the resource domain set is
$AD(M_r^v)$. A Root-accepted record contains at least:

\[
  \langle did_r, v, H(M_r^v), AD(M_r^v), status, t, \sigma_\rho \rangle .
\]

The record binds resource DID, version, representation hash, authorized
domains, lifecycle status, timestamp or logical time, and Root signature. The
exact serialization may evolve, but the binding requirement is stable.
Resource acceptance is Root anchored, while infrastructure eligibility is
governance-backed and Root-credentialed.

An identity representation is admissible to a Discovery node $d$ only if it is
well formed, anchored by Root, current, and authorized for $d$:

\[
\begin{aligned}
Admissible(d,M) ={}& WellFormed(M) \land Anchored(M)\\
&{}\land Current(M) \land Scope(d,M).
\end{aligned}
\]

This predicate separates business search from trust admission. A query may
match many metadata fields, but only admissible candidates should be exposed.

\begin{table}[t]
\centering
\caption{Stable Technical Terms}
\label{tab:stable-terms}
\scriptsize
\begin{tabular}{@{}p{0.28\columnwidth}p{0.62\columnwidth}@{}}
\toprule
Term & Meaning in this paper \\
\midrule
Root Node & trust-domain service for acceptance, proof, publication, and infrastructure VCs \\
Registrar node & governed resource onboarding and Root submission node \\
Discovery node & package synchronization, indexing, query, and signed response node \\
Trust Indexer & component that converts governance events into runtime-readable authorization state \\
Public visibility surface & human-facing aggregation of metrics or summaries, not a protocol authority \\
Authorized domains & governance scope used for Registrar and Discovery authorization \\
Capability tags & semantic search and matching signals, not authorization scope \\
\bottomrule
\end{tabular}
\end{table}

Table~\ref{tab:stable-terms} defines the technical terms used throughout this
paper. Root Node and Root service denote a protocol role independently of the
service instances that implement it. Public portals and network transparency
surfaces expose runtime state without introducing additional protocol entities.

\section{Role Architecture}

\begin{figure*}[!ht]
\centering
\includegraphics[width=0.9\linewidth]{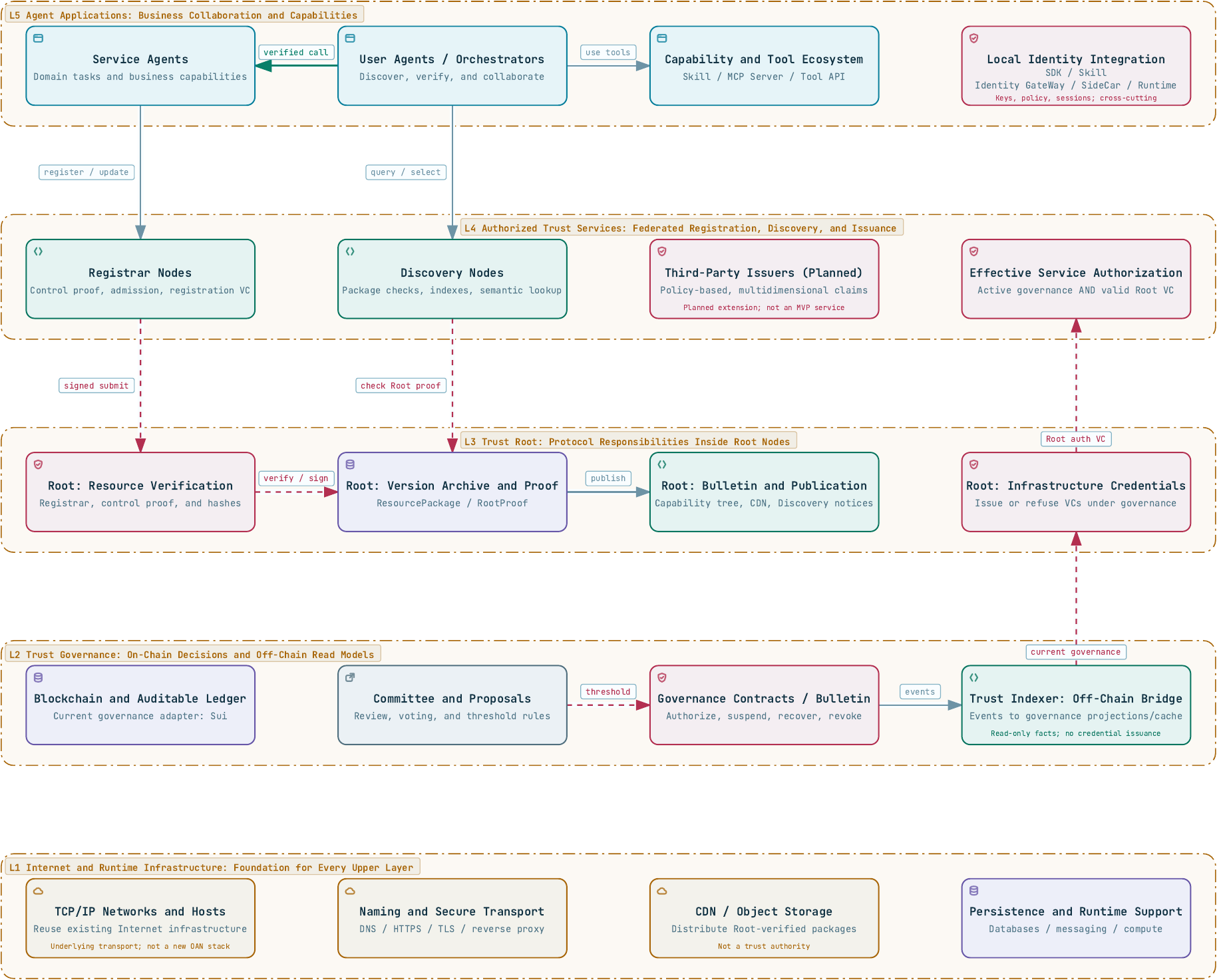}
\caption{Five-layer OAN architecture from Internet infrastructure and on-chain
governance to Root responsibilities, authorized trust services, and Agent
applications. Third-party credential issuance is shown as a planned extension.}
\label{fig:yp-five-layer-architecture}
\end{figure*}

The technical roles are intentionally separated. Figure~\ref{fig:yp-five-layer-architecture}
places them across the Internet foundation, trust governance, Root enforcement,
authorized services, and Agent application layers. These architectural layers
describe responsibilities rather than the L0--L3 identity-integration depths.

\subsection{Root}

Root owns accepted-version state, authorization-domain enforcement metadata,
semantic tag tree governance, bulletin facts, infrastructure authorization VC
issuance, and verified package generation. It consumes infrastructure lifecycle
state from the on-chain governance layer and verifies Registrar authorization,
signed upstream envelopes, registration credentials, DID structures, document
hashes, capability metadata, and status transitions. Root should be the source
of truth for current accepted resource identity versions, but not the sole
governance source for infrastructure-node lifecycle legitimacy.

Root is not the business data plane. It should not need to observe every
Agent-to-Agent call. This reduces operational load and avoids turning the trust
anchor into a universal traffic proxy.

Root can be modeled as three coordinated hubs. As the trust authorization hub,
it accepts resource versions, issues infrastructure credentials, and signs
protocol evidence. As the data-distribution hub, it turns accepted identity
state into Root-verified packages and publication cursors. As the
semantic-governance hub, it governs the semantic tag tree used by Registrar for
tag selection and recommendation and by Discovery for semantic matching,
ranking, and explanation. The last role governs vocabulary and versions; it
does not require Root to execute user-facing semantic queries.

In this governance model, Root acts as the policy enforcement point and protocol
credential issuer for infrastructure nodes. A governance state transition does
not by itself create a usable protocol credential. Root issues the
infrastructure authorization VC after it observes active governance state and
verifies DID-control and operational materials. Conversely, Root should stop
serving or accepting later requests from an infrastructure subject whose latest
governance state is inactive, even if the subject still holds an old VC.

Root-verified packages and governance-derived authorization state provide
consistent evidence for machine verification and human inspection. Public
interfaces expose this evidence without becoming alternative sources of trust.

\subsection{Registrar}

Registrar owns onboarding assistance, draft workflow, product-type declaration,
semantic description collection, capability-tag suggestion, subject-control
challenge handling, registration credential issuance, and Root submission. It
is an operational gateway between resource providers and Root. Registrar can
improve usability and review quality, but it does not unilaterally decide final
Root acceptance.

\subsection{Discovery}

Discovery owns package synchronization, Root proof verification, bulletin
checking, authorization-domain filtering, local indexing, query handling, and
response signing. Discovery is allowed to optimize search and ranking, but its
first obligation is to avoid exposing non-admissible identity representations.
In the on-chain governance model, Discovery consumes authorization events
through an indexer or local cache; it does not need to embed proposal or voting
logic in the default service runtime.

\subsection{CDN}

CDN owns distribution of Root-verified packages and manifests. CDN improves
availability and scalability but is not a trust authority. Consumers should
verify package hashes, Root proof, metadata, and bulletin consistency.

\subsection{Resource Provider and Resource Consumer}

The Resource Provider controls a registered Agent Service, Skill, MCP Server,
or Tool/API resource. The Resource Consumer is the relying Agent, application,
developer tool, or human-facing client. Their trust relationship begins with
Discovery verification and continues through signed invocation, credential
checks, or artifact access according to the product type. OAN does not dictate
the complete application conversation after the trust guard passes.

\begin{figure*}[!ht]
\centering
\includegraphics[width=0.8\linewidth]{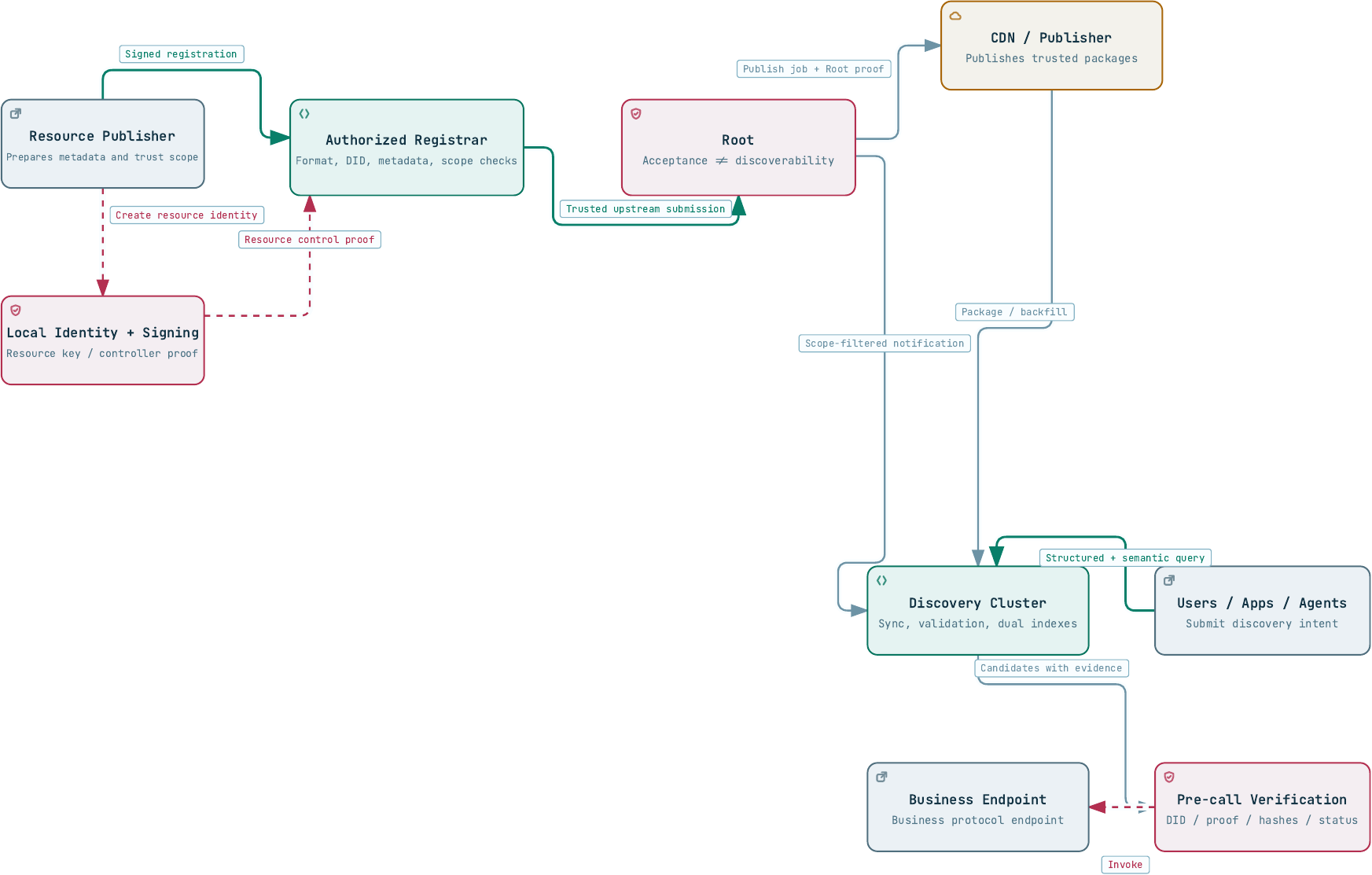}
\caption{Resource identity lifecycle from local preparation and Registrar admission through Root publication, Discovery, and verified invocation.}
\label{fig:yellowpaper-lifecycle}
\end{figure*}

Figure~\ref{fig:yellowpaper-lifecycle} connects these roles through the
registration, publication, discovery, and verified-invocation sequence.

\section{Protocol Objects}

\begin{figure*}[!ht]
\centering
\includegraphics[width=0.8\linewidth]{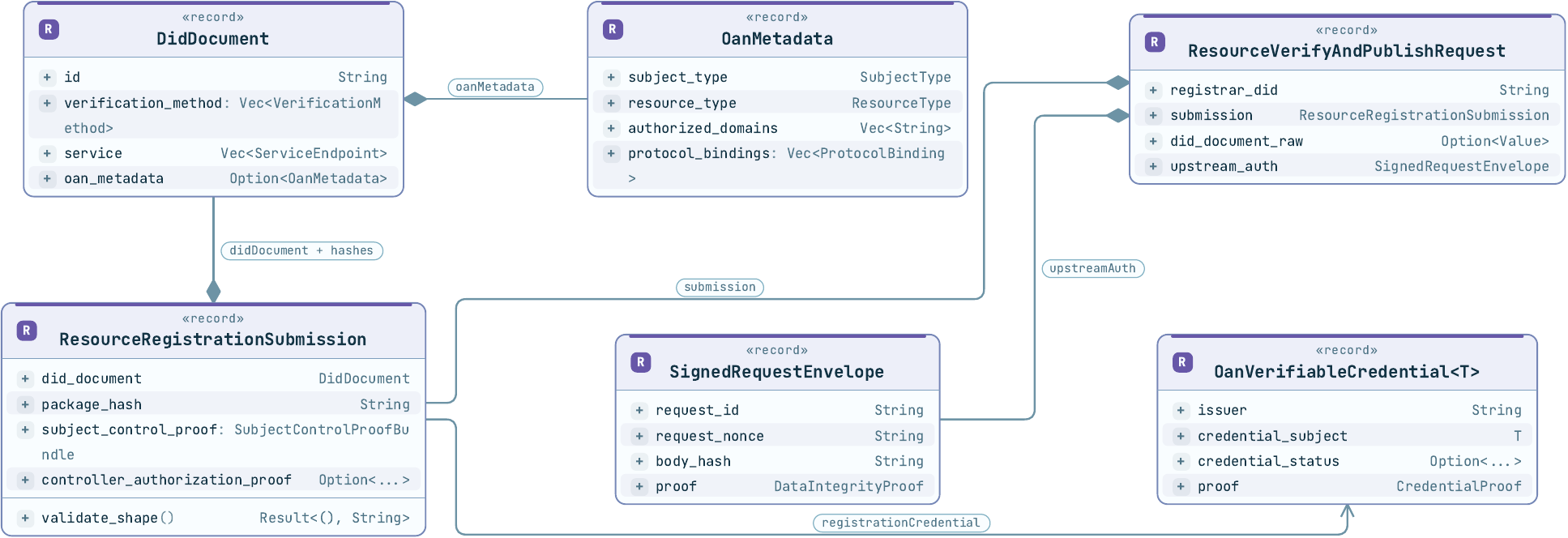}
\caption{Identity and registration protocol objects: local OAN Identity, public DID Document, control proof, and Registrar-issued registration credential.}
\label{fig:yp-identity-registration-objects}
\end{figure*}

Figure~\ref{fig:yp-identity-registration-objects} maps the DID Document,
metadata, proof-bearing submission objects, and Registrar-issued credential.
The following subsections distinguish their contents and verification roles.

\subsection{\texttt{did:oan} Identity Representation}

The identity representation is centered on a \texttt{did:oan} DID Document
compatible with DID Document concepts. A \texttt{did:oan} identifier has the
form:

\begin{center}
\texttt{did:oan:<routing-code>:<suffix-code>}
\end{center}

The \texttt{routing-code} is a method-specific routing or issuance-origin hint,
while the \texttt{suffix-code} carries subject uniqueness. In the OAN
reference trust path, infrastructure-node DIDs derive their routing relationship
from the Root DID, and resource DIDs derive their routing relationship from the
Registrar that admits the resource. This routing relationship is an admission
and consistency check. It is not a proof of control, credential validity, or
governance state by itself.

An OAN DID Document contains identifier, controller relationship, verification
methods, authentication and assertion relationships, service endpoints or
artifact references, OAN metadata, and a Data Integrity proof. The default
verification method is \texttt{<did>\#key-1}. DID Document and VC proofs use the
\texttt{Ed25519Signature2020} Data Integrity style with a multibase
\texttt{proofValue}. Digest algorithms declared in resource-package or metadata
fields describe content hashes; they do not select the proof suite.

The DID Document contains resource product type, service endpoints or artifact
references, semantic description fields, capability tags, typed
\texttt{authorizedDomains}, credential references, OAN metadata, and version
metadata. The \texttt{did:oan} method provides the resource identifier used
throughout registration, Root acceptance, and Discovery.

The identity representation should include enough material to support:

\begin{itemize}[leftmargin=*]
  \item key discovery and signature verification;
  \item endpoint or artifact-reference discovery for supported product types;
  \item authorized-domain validation and Discovery routing;
  \item semantic search over labels, descriptions, examples, and use cases;
  \item credential binding;
  \item version and status tracking;
  \item Root package generation and verification.
\end{itemize}

The core product types are Agent Service, Skill, MCP Server, and Tool/API.
Agent Service, MCP Server, and Tool/API resources normally provide network
endpoints. A Skill may instead provide an external artifact uniform resource
identifier (URI) and enough
semantic metadata for discovery. OAN Discovery indexes the DID Document and
Root-verified package metadata; it does not need to host external artifacts.

The DID string should identify the resource subject, not encode mutable
deployment facts. Endpoint URLs, artifact locations, schema references,
descriptions, capability tags, authorization domains, status, and version
metadata belong in the DID Document, product metadata, or Root-verified package.
This separation keeps the identifier stable while allowing endpoints, keys,
resource descriptions, and package references to evolve through normal
versioning and Root acceptance.

\subsection{OAN Identity and Controller Relationships}

OAN distinguishes public identity documents from local identity material. An
OAN Identity is local private material that binds a \texttt{did:oan} identifier,
the default verification method, a DID Document, public key material, and the
corresponding private key material. It is used to create DID Document proofs,
request proofs, VC proofs, or application login challenge signatures. It is not
published as a DID Document and must not be treated as a credential by itself.

This distinction supports three controller cases. In a self-controlled
resource, the DID Document's \texttt{controller} is the resource DID itself.
A dedicated controller can also be registered as a normal OAN resource, often
using a controller-oriented subject and resource type. A third resource may be
controlled by an external controller DID, in which case a controller
authorization proof binds the resource DID, controller DID, Registrar DID,
document hashes, verification method, purpose, validity window, and nonce.
Registrar and Root both verify this proof when it is required. A self-controlled
resource does not need an external controller authorization proof.

For an existing resource, authorization is evaluated against the controller in
the current Root-accepted version, not only against the controller declared by
the proposed document. If version $v$ is controlled by $A$ and version $v+1$
declares $B$, $A$ authorizes the transition; $B$ authorizes subsequent updates.
The current controller may designate its successor without a separate
acceptance signature from that successor. Root commits the accepted version and
controller state atomically, so concurrent updates from the same predecessor
cannot both become current.

\subsection{External Identifier Compatibility}

OAN identifiers remain the primary identifiers inside the OAN trust path, but a
DID Document may also declare external identifiers. An
\texttt{externalIdentifiers} item contains an \texttt{id}, which is the native
identifier in an external platform or identification system, and an optional
\texttt{resolutionServiceEndpoint}, which gives the resolution context. The
\texttt{id} is a metadata string, not the DID Document's top-level
\texttt{id}, not a DID URL, and not a JSON-LD node identifier. Values such as
repository paths, platform-local handles, industrial identifiers, or object
identifiers can therefore be represented without rewriting them into
\texttt{did:oan} syntax.

External identifiers improve interoperability with existing platforms and
industry identifier systems, but they do not change DID control, Root
acceptance, authorized-domain scope, or Discovery trust. They are
signed declarations bound to the resource DID Document. Stronger assurance of
external ownership or endorsement requires evidence from the relevant external
system or an authorized credential issuer. Table~\ref{tab:resource-types}
summarizes the OAN material and native responsibility of each initial type.

\begin{table}[t]
\centering
\scriptsize
\caption{Initial OAN Resource Types}
\label{tab:resource-types}
\resizebox{\columnwidth}{!}{%
\begin{tabular}{@{}llll@{}}
\toprule
Type & Native object & Required OAN material & Native responsibility \\
\midrule
Agent Service & callable Agent & endpoint, keys, claims & task protocol \\
Skill & skill artifact & artifact URI/hash, semantics & package format \\
MCP Server & MCP endpoint & endpoint, protocol hints & MCP behavior \\
Tool/API & API endpoint & endpoint/schema reference & API semantics \\
\bottomrule
\end{tabular}
}
\end{table}

\subsection{Product-Type Metadata}

Product-type metadata declares how a resource should be interpreted without
forcing every resource into the Service Agent shape. The minimal model should
include product type, version, semantic labels, natural-language description,
example use cases, authorized domains, endpoint descriptors or artifact
references, and protocol hints. Product-native schemas remain outside the OAN
core. For example, MCP tool listings, OpenAPI documents, or Skill package files
can be referenced by URI and verified by hash when needed.

The product-type field is not an authorization shortcut. A resource marked as
an MCP Server or Tool/API still needs normal Registrar onboarding, Root
acceptance, package verification, and Discovery authorization checks before it
is exposed. Its \texttt{authorizedDomains} must remain explicit and must not be
inferred from product type, endpoint hostnames, or capability tags.

Product-type metadata should separate \emph{trust fields} from \emph{search
fields}. Trust fields include DID, verification methods, resource type,
version, status, endpoint or artifact-reference hash, credential references,
and critical extension markers. Search fields include labels, descriptions,
examples, natural-language use cases, local tags, and ranking hints. A verifier
may use search fields to rank results, but it must not let search fields change
authorization, current-version, or signature decisions.

Implementations should preserve product-native differences without giving them
different trust semantics. An MCP Server can keep MCP descriptors, a Skill can
keep its package format, and a Tool/API can keep an OpenAPI or domain schema.
OAN only requires that the native material be referenced, hashed, typed, and
bound to the same DID, Registrar, Root, Discovery, and lifecycle checks.

\subsection{ResourcePackage as the Registry Object}

The ResourcePackage is the registry object that makes OAN resource-first. It is
not just an endpoint listing or a marketplace item. It binds:

\begin{itemize}[leftmargin=*]
  \item the \texttt{did:oan} resource identifier and DID Document hash;
  \item resource type, version, status, authorized domains, and semantic
  metadata;
  \item endpoint descriptors or external artifact references;
  \item Registrar credential evidence and subject-control binding;
  \item Root acceptance proof, publication cursor, and package hash;
  \item optional schema records, such as an external Agent-schema record, when a
  deployment chooses to use them.
\end{itemize}

This package can carry an MCP Server entry, a Skill identity, an Agent Service,
or a Tool/API descriptor through the same lifecycle. Product-native metadata
can remain in its native format, but it is placed inside or referenced by a
Root-verified envelope. This is the main technical difference between OAN and a
narrow registry that only stores tool endpoints or Skill artifacts.

External artifact references should be explicit. A ResourcePackage that points
to a Skill file, OpenAPI document, MCP descriptor, or other artifact should
include at least URI, media type, digest algorithm, digest value, expected
version, and optional retrieval policy. OAN does not guarantee that the remote
artifact host is trustworthy; it guarantees that the reference and digest were
part of the Root-verified package. Consumers must still verify the digest after
retrieval.

\subsection{Capability Metadata}

Capability metadata has a canonical component and an extensible component. The
canonical component maps to the governed authorization-domain vocabulary and
the governed semantic tag tree. The extensible component allows
application-specific capability tags, search hints, labels, or descriptions.
Canonical authorized domains, not custom tags, determine Registrar admission
and Discovery exposure. Governed semantic tags and custom tags may refine
registration suggestions, search, and ranking within an authorized set, but
they must not grant exposure outside authorized domains.

\subsection{Registration Credential}

The resource registration credential binds:

\begin{itemize}[leftmargin=*]
  \item issuer Registrar DID;
  \item subject resource DID;
  \item credential type and claims;
  \item registration binding, including DID-control evidence;
  \item issuance time, validity, and proof.
\end{itemize}

Root verifies issuer authorization, credential proof, subject binding,
credential status, product-type metadata, and validity before accepting a
submitted representation.
The credential attests only the onboarding claims defined by its profile. A
binding to a verified legal person or enterprise requires an explicit
Registrar policy and corresponding credential claims; it is not implied by
the generic registration credential.

Infrastructure authorization credentials, resource registration credentials,
and third-party business credentials are distinct VC types. They can share the
same outer VC and Data Integrity proof model, but their issuers, subjects, and
status services differ. A VC's \texttt{credentialStatus} points to the issuer's
status service. A status value embedded in the VC is a signing-time or
query-time projection, not an automatically updating state. The Trust Indexer
and other governance indexers
or other independent evidence sources may be used for higher-assurance checks,
but they do not replace the issuer status interface as the general VC status
model.

\subsection{Subject-Control Challenge}

Subject-control proof prevents a Registrar from issuing a credential for a DID
that the applicant cannot control. The challenge should bind draft identifier,
subject DID, Registrar DID, expiration time, and nonce. The Agent signs the
challenge using an assertion-capable verification method from its DID document.
Registrar verifies the signature and stores evidence or a verifiable summary.

\subsection{Signed Upstream Envelope}

Privileged infrastructure writes, especially Registrar-to-Root submissions, use
a signed upstream request envelope. The envelope covers sender DID, target
audience, method, path, body hash, timestamp, nonce, purpose, and protocol
version. Root verifies the sender's current authorization and checks replay
state before processing the body.

\subsection{Root-Verified Package}

The package consumed by CDN and Discovery includes the DID Document, metadata,
hashes, Root proof, publication cursor, subject version, status, and capability
information. Discovery does not trust the CDN path alone; it recomputes hashes
and verifies Root proof and bulletin consistency.

\begin{figure*}[!ht]
\centering
\includegraphics[width=0.9\linewidth]{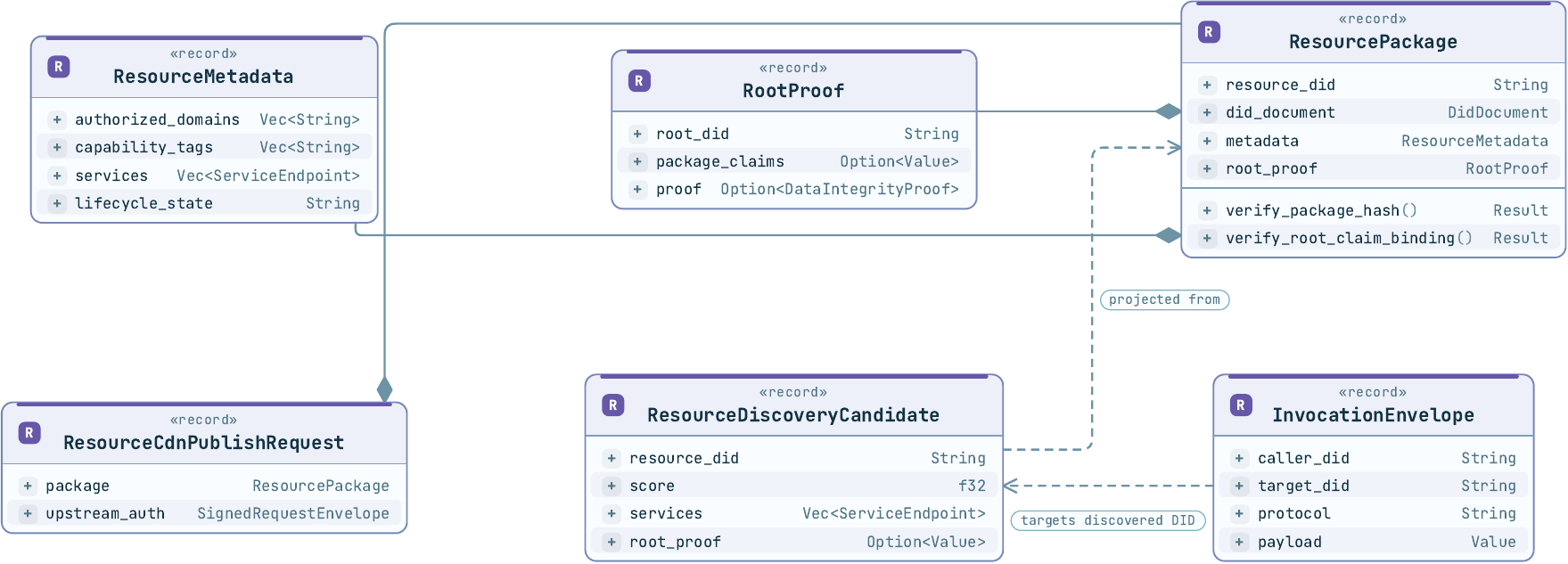}
\caption{Root-verified package and downstream discovery objects used to bind publication evidence to a later invocation.}
\label{fig:yp-package-discovery-objects}
\end{figure*}

Figure~\ref{fig:yp-package-discovery-objects} shows how publication evidence
is carried into downstream Discovery responses and invocation context.

\subsection{Discovery Response}

Discovery responses are signed by the Discovery node and include returned
candidates, provenance metadata, timestamp, query context, and proof. A relying
User Agent can verify that the response came from a currently authorized
Discovery node and that selected candidates trace back to Root-verified
packages.

\subsection{Root Identity Directory}

Some applications need to verify that a controller or resource identity can
authenticate with an OAN Identity. For this purpose, Root may expose an
identity directory that projects Root-accepted identities and their verified
public keys. The directory is a public-key discovery aid for application
authentication. It does not replace DID resolution, VC verification, resource
authorization, or the controller proof used during resource registration.

An identity directory item minimally contains the subject DID, status, verified
public keys, an update time, and a revision indicator. Applications authenticate
the holder by checking the directory entry and verifying a fresh challenge
signature produced with the holder's OAN Identity. Public keys supplied by the
holder require validation against Root-accepted evidence before use.

The directory is also not a cross-resource control list. A controller that
appears in the directory may authenticate to an application, but resource
registration still requires the Registrar and Root to validate the applicable
resource DID, controller relationship, controller authorization proof, DID
Document hashes, authorized-domain scope, and replay protections. This
distinction keeps login public-key discovery separate from registration
authorization.

\subsection{Trusted Invocation Envelope}

Before invoking a callable resource such as an Agent Service, the Resource
Consumer sends caller DID material, credentials, nonce, timestamp, target DID,
request body hash, discovery provenance, and request proof. The Resource
Provider checks caller identity, credential subject, target binding, body hash,
signature, nonce uniqueness, and timestamp freshness. Non-callable resources
such as Skill artifacts can still use the same discovery and package evidence
before the consumer follows an external artifact reference.

\begin{table}[t]
\centering
\scriptsize
\caption{Core Object Responsibilities}
\label{tab:object-responsibilities}
\resizebox{\columnwidth}{!}{%
\begin{tabular}{@{}llll@{}}
\toprule
Object & Producer & Consumer & Purpose \\
\midrule
Identity doc & Resource provider & Registrar, Root & subject keys \\
Reg credential & Registrar & Root & onboarding proof \\
Bulletin event & gov. client & infra nodes & lifecycle fact \\
Verified package & Root & CDN, Discovery & accepted version \\
Discovery response & Discovery & Resource consumer & query provenance \\
Invocation envelope & Resource consumer & callable provider & request binding \\
Signed response & callable provider & Resource consumer & response origin \\
\bottomrule
\end{tabular}
}
\end{table}

Table~\ref{tab:object-responsibilities} associates each object with its producer,
consumer, and verification purpose. These responsibilities remain distinct.
For example, a Discovery response does not create Root
acceptance; it only presents signed query provenance over already verified
packages. A registration credential does not make an Agent discoverable; it
only supports Root admission.

\section{On-Chain Governance Layer}

The OAN governance profile uses an on-chain governance layer as the
event source for infrastructure-node lifecycle state. Related self-sovereign identity (SSI) systems have
also explored distributed trust anchors and lifecycle
coordination~\cite{gruner2023ssi,glockler2023iam,popa2023chaindiscipline}.
The scope is intentionally narrow. It manages lifecycle state for Registrar,
Discovery, and future third-party VC issuer nodes. It does not manage resource
registration records, full identity documents, resource metadata updates, Root
key rotation, or CDN service coordination.

This layer separates infrastructure governance from service operation.
A committee or other threshold-governed
operator set can decide infrastructure lifecycle state, while Root enforces the
result and signs the protocol artifacts that ordinary OAN services understand.
The governance layer publishes federated trust decisions, while Root enforces
those decisions and issues protocol credentials.

\subsection{Lifecycle Scope}

The governance layer records and emits events for:

\begin{itemize}[leftmargin=*]
  \item Registrar authorization, authorized-domain updates, suspension,
  recovery, and revocation;
  \item Discovery authorization, authorized-domain updates, suspension,
  recovery, and revocation;
  \item third-party VC issuer authorization and revocation.
\end{itemize}

The event stream replaces repeated status checks against Root for these
infrastructure lifecycle facts. Registrar, Discovery, and VC issuer services
consume indexed governance events, rebuild local authorization state, and
enforce the resulting state in their own service logic. Root also consumes the
same indexed state so that online VC issuance, Registrar submissions, Discovery
notification, and future VC-issuer authorization can fail closed when the
latest governance state is inactive or too stale.

Authorized-domain payloads are typed governance data, not opaque extension
metadata. The all-domain value is represented by \texttt{["*"]}; an empty list
means no domain authorization. A parent domain covers descendants by dot-prefix
hierarchy. The taxonomy draws on Agent capability classification work,
including Open Agentic Schema Framework (OASF)-inspired
categories~\cite{agntcy_oasf}, while its identifiers and evolution are governed
within OAN.

\subsection{Governance Operation Boundary}

In a governed deployment, governance operations are performed by designated
governance clients rather than by ordinary Registrar, Discovery, or VC issuer
service nodes. These clients create proposals, submit votes, refresh proposal
status, and manage participant lifecycle state.

Registrar, Discovery, and VC issuer service nodes are event consumers by
default. Their runtime should include event subscription or polling and local
authorization-cache maintenance, but it should not need proposal, vote, or
member-management code. This separation keeps service nodes simpler and keeps
governance private-key handling outside ordinary service runtimes.

\subsection{DID and VC Boundary}

The on-chain governance layer does not store, parse, validate, or issue DID
Documents or Verifiable Credentials. It also does not model the internal schema
of \texttt{did:oan}, resource identifiers, credential claims, credential
proofs, resource metadata, or capability descriptions.

Fields such as subject identifier, policy hash, metadata hash, effective time,
expiration time, and status are only lifecycle references. DID and VC semantics
remain off-chain OAN protocol responsibilities implemented by Root, Registrar,
Discovery, VC issuer services, and SDK verification logic.

Effective infrastructure authorization requires both an active governance state
and a valid Root-issued infrastructure authorization VC. If governance state is
inactive, revoked, unknown, or too stale to trust, a previously issued VC is not
sufficient for new service. If governance state is active but Root has not
issued a VC, the node has lifecycle approval but is not yet authorized at the
OAN protocol layer. This distinction is important for interoperability: service
nodes can validate ordinary protocol credentials, while infrastructure
operators and indexers can validate the fresher governance state.
Table~\ref{tab:blockchain-bulletin-boundary} separates governance-layer facts
from the resource and credential operations that remain off chain.

\begin{table}[t]
\centering
\caption{On-Chain Governance Boundary}
\label{tab:blockchain-bulletin-boundary}
\begin{tabular}{@{}ll@{}}
\toprule
In governance-layer scope & Outside governance-layer scope \\
\midrule
Registrar lifecycle & Resource DID Document storage \\
Discovery lifecycle & VC issuance or validation \\
VC issuer lifecycle & Resource metadata parsing \\
Discovery domain updates & Resource metadata and package content \\
Lifecycle events & Root key and CDN coordination \\
\bottomrule
\end{tabular}
\end{table}

\section{Lifecycle State Machine}

OAN treats identity as stateful. A typical resource identity moves through the
states shown in Table~\ref{tab:lifecycle-states}.

\begin{table}[t]
\centering
\caption{Resource Identity Lifecycle States}
\label{tab:lifecycle-states}
\scriptsize
\begin{tabular}{@{}p{0.26\columnwidth}p{0.22\columnwidth}p{0.42\columnwidth}@{}}
\toprule
State & Owner & Meaning \\
\midrule
draft & Registrar & identity under preparation \\
control-verified & Registrar & DID control proof passed \\
credentialed & Registrar & registration credential issued \\
submitted & Registrar/Root & Root submission in progress \\
accepted & Root & version accepted by Root \\
published & Root/CDN & package ready for sync \\
synced & Discovery & package received and verified \\
structured-indexed & Discovery & structured index updated \\
semantic-indexed & Discovery & semantic index updated or marked recoverable \\
discoverable & Discovery & admissible for query within scope \\
superseded & Root & replaced by newer version \\
revoked & Root & no longer admissible \\
\bottomrule
\end{tabular}
\end{table}

Create and update should use the same full-document path. Root should verify
the complete new representation rather than applying ambiguous partial patches.
This makes current-version verification easier and reduces risk that old fields
remain accidentally authoritative.
The table also separates acceptance from visibility. \texttt{accepted} does not
mean immediately discoverable. A resource may be accepted by Root, published to
the distribution layer, synchronized by Discovery, indexed structurally,
indexed semantically, and then exposed as a query candidate. These steps may
complete at different times, but they should leave observable state so that
operators can distinguish lag from rejection.

The lifecycle is not a one-time publish action. Implementations should handle
update, suspension, recovery, revocation, and republish as explicit state
changes with verifiable causes. A newer accepted version should supersede older
versions for normal Discovery, a suspended or revoked resource should not remain
visible after synchronization, and a republished resource should re-enter
Discovery through the same package-verification and authorization checks rather
than bypassing them as a cache update.

\section{Lifecycle Flow}

The standard OAN lifecycle is:

\begin{enumerate}[leftmargin=*]
  \item Resource provider prepares a \texttt{did:oan} identity representation
  with explicit \texttt{authorizedDomains}.
  \item Registrar assists draft creation, authorized-domain selection, and
  capability-tag selection.
  \item Resource provider proves control of its DID key.
  \item Registrar verifies that the final resource domains are covered by its
  own authorized domains and issues a resource registration credential.
  \item Registrar submits the complete representation to Root with a signed
  upstream envelope.
  \item Root verifies Registrar authorization, envelope proof, credential
  proof, subject binding, DID structure, and capability metadata.
  \item Root archives the accepted version and creates a Root-verified package.
  \item CDN distributes the package.
  \item Root targets Discovery notification by resource
  \texttt{authorizedDomains} and Discovery authorized-domain coverage.
  \item Discovery synchronizes packages, verifies Root proof and bulletin
  facts, applies authorized-domain filtering, and indexes eligible Agents.
  \item User Agent queries Discovery and verifies the signed response.
  \item User Agent invokes Service Agent with a trusted invocation envelope.
  \item Service Agent verifies the envelope and returns a signed response.
\end{enumerate}

\begin{figure*}[!ht]
\centering
\includegraphics[width=0.8\linewidth]{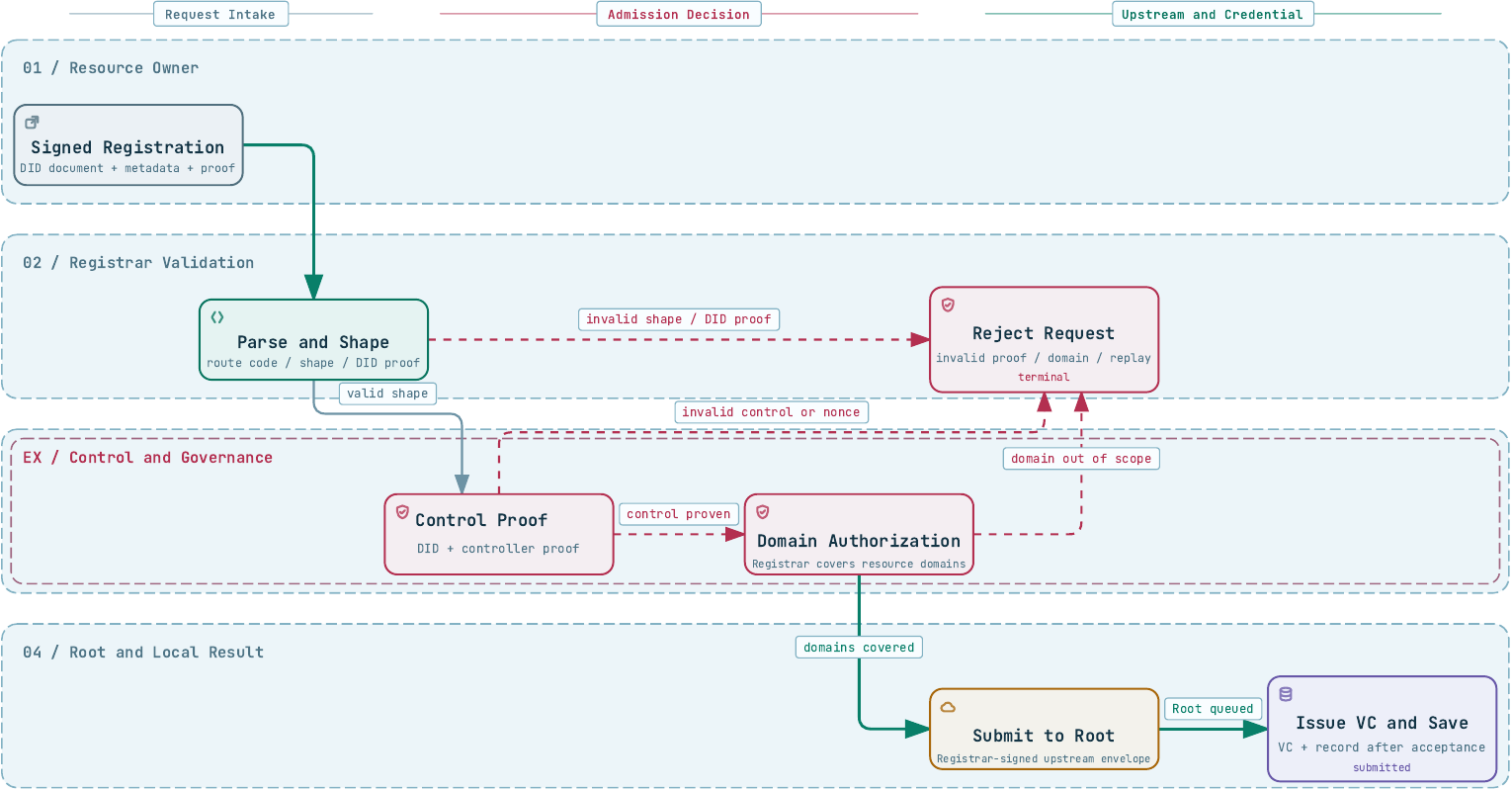}
\caption{Registrar admission checks for resource identity, subject and controller proofs, authorized scope, and registration credential issuance.}
\label{fig:yp-registrar-admission}
\end{figure*}

Figure~\ref{fig:yp-registrar-admission} expands the Registrar admission stage:
subject and controller proofs and authorized scope are checked before credential
issuance and subsequent Root verification.

\section{Root Verification Requirements}

Root verification should be deterministic and auditable. At minimum, Root
should verify:

\begin{itemize}[leftmargin=*]
  \item signed upstream envelope integrity;
  \item sender Registrar DID and authorization status;
  \item timestamp freshness and nonce uniqueness;
  \item request body hash binding;
  \item registration credential proof and issuer binding;
  \item subject DID binding between credential and document;
  \item DID document structure and verification methods;
  \item resource \texttt{authorizedDomains} against the Registrar's authorized
  domains;
  \item capability tags against the governed taxonomy as search metadata, not
  authorization evidence;
  \item semantic tag references against the current or compatible governed tag
  tree version;
  \item version and status transition validity;
  \item for an update, authorization by the controller in the current
  Root-accepted version and an atomic transition of version and controller
  state.
\end{itemize}

Root should return an explicit rejection when a required check fails.
Downstream Discovery nodes and relying consumers use Root acceptance as
evidence that the required admission checks have completed; partial
verification must not be reported as acceptance.

\section{Discovery Authorization}

OAN uses authorized domains to control which Registrar may admit a resource and
which Discovery node may expose it. Let $C(X,Y)$ mean that authorized-domain
set $X$ covers every domain in set $Y$, where \texttt{["*"]} covers all
domains and a parent domain covers descendants by hierarchy. Let $Auth(rg)$ be
the authorized domain set for Registrar $rg$, $Auth(d)$ the authorized domain
set for Discovery node $d$, and $AD(M)$ the resource authorized-domain set in
identity representation $M$.

\[
\operatorname{RegScope}(rg,M)=
\begin{cases}
true, & C(Auth(rg), AD(M)),\\
false, & otherwise.
\end{cases}
\]

\[
DiscoveryScope(d,M)=
\begin{cases}
true, & C(Auth(d), AD(M)),\\
false, & otherwise.
\end{cases}
\]

Custom tags are allowed, but they cannot expand the governance domain. They
are used only after coarse authorization for local refinement, ranking, or
search. A resource with empty \texttt{authorizedDomains} is not admissible for
normal registration unless a local onboarding policy computes an explicit
covered subset before Root submission.

This rule is intentionally strict because capability labels are often noisy,
user-facing, multilingual, or generated by operators. A label such as
\texttt{finance}, \texttt{health}, or \texttt{coding} may be useful for semantic
matching, but it cannot prove that a Registrar is allowed to admit the resource
or that a Discovery node is allowed to expose it. Authorization must come from
the governed domain vocabulary and the corresponding node authorization state.
The governed semantic tag tree complements this authorization vocabulary. It
helps Registrar nodes recommend consistent tags during registration and helps
Discovery nodes map user intent to comparable resource metadata during semantic
discovery, but it is not an authorization grant.

\subsection{Extension Safety Rules}

Extensions are useful for local search quality and domain integration, but they
must not weaken authorization. The following rules should hold for custom tags,
plugin metadata, semantic embeddings, ranking features, and domain-specific
attributes:

\begin{itemize}[leftmargin=*]
  \item an extension must not cause Registrar or Discovery to admit or expose
  an identity outside the applicable \texttt{authorizedDomains};
  \item an extension must not override Root acceptance, current-version state,
  revocation state, or package verification results;
  \item ranking and semantic matching must run only after admissibility checks;
  \item unknown non-critical extension fields may be ignored, while unknown
  critical fields must cause rejection by verifiers that do not support them;
  \item plugin output should be auditable enough for an operator to explain why
  an identity was indexed, rejected, ranked, or returned.
\end{itemize}

\subsection{Synchronization Rules}

Discovery synchronization should be incremental. A publication cursor or
watermark allows Discovery to fetch new packages without repeatedly scanning
the entire Root archive. Discovery should verify each package before indexing
and should track rejected packages with reasons. A package that fails Root
proof, hash, bulletin, status, or authorization checks should not enter the
query index.

\subsection{Query Rules}

Discovery query can use keywords, capability tags, endpoint metadata, local
indexes, or ranking signals. However, query matching should never bypass the
admissibility predicate. The proper order is: verify package, check current
state, enforce Discovery authorization, index admissible representation, then
apply query matching and ranking.

Semantic discovery is an implementation profile above this trust filter. GRAIL
studies deep-granularity hybrid resonance and small language model
(SLM)-enhanced indexing for
real-time Agent discovery~\cite{xu2026grail}; OAN can use this kind of
semantic indexing inside Discovery nodes to map natural-language user intent to
verified resource metadata. The invariant is that semantic matching ranks or
filters only candidates that are already admissible under Root proof,
current-version state, and Discovery authorization.

When a governed semantic tag tree is available, Discovery should bind its
semantic index to an explicit tag-tree version or compatibility range. A
Discovery node may use embeddings, local synonyms, examples, and ranking
features, but those local features should be traceable back to governed tags
when they affect cross-node resource comparability.

A semantic result should therefore be inspectable. Discovery may explain why a
candidate matched, return highlighted metadata, or expose the underlying DID
Document and package evidence for client verification. These inspection fields
improve user trust and debugging, but they do not replace signed Discovery
responses or Root package verification.

\begin{figure*}[!ht]
\centering
\includegraphics[width=0.8\linewidth]{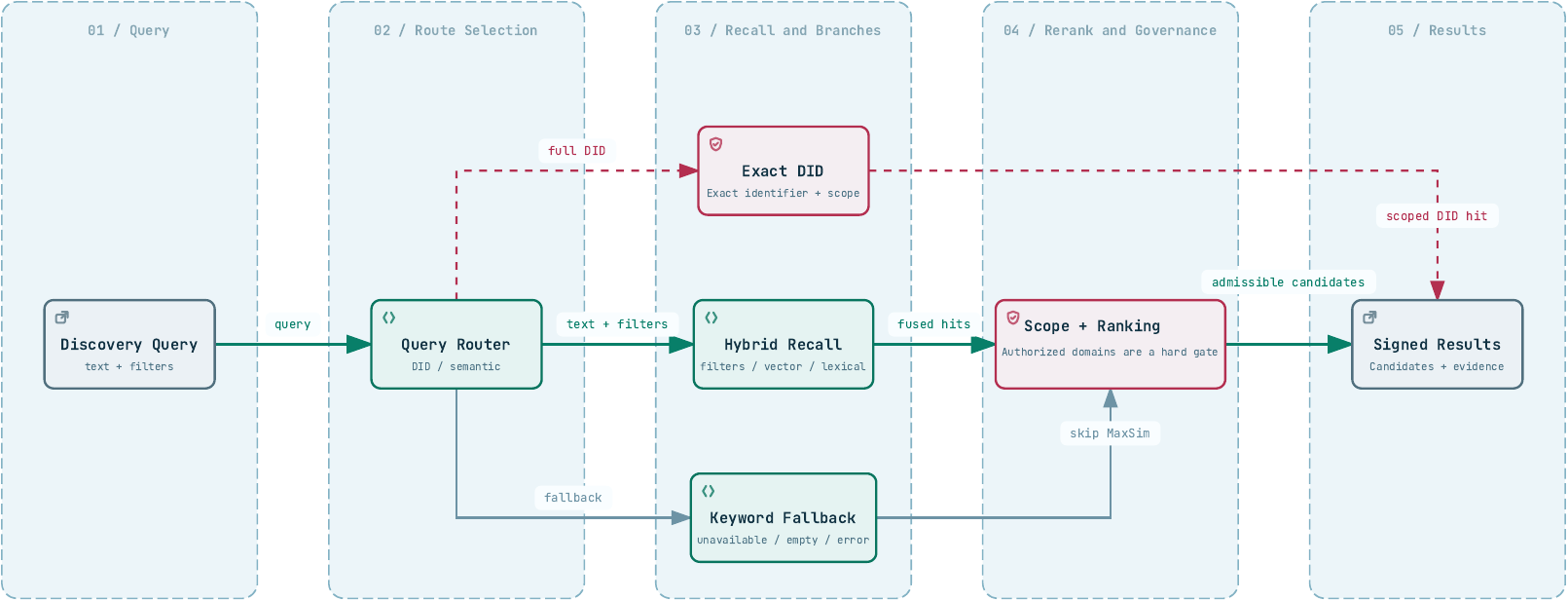}
\caption{Discovery query paths: exact DID lookup and semantic recall operate only within the authorized and admissible candidate set.}
\label{fig:yp-discovery-query}
\end{figure*}

Figure~\ref{fig:yp-discovery-query} contrasts exact DID lookup with semantic
recall while preserving the same authorized and admissible candidate boundary.

\section{Trusted Invocation}

Trusted invocation binds a discovered resource identity to a protocol-specific
request. The base envelope defines the checks a Service Agent applies before
passing the request to its business handler. Protocol-specific message
protection beyond this initial request is defined by the integration profiles
described below.

A Service Agent should verify:

\begin{itemize}[leftmargin=*]
  \item caller DID document and verification method;
  \item caller credential proof and subject binding;
  \item Discovery response proof and freshness, when included;
  \item selected-service provenance against Root-verified package data;
  \item target DID equals the invoked Service Agent DID;
  \item timestamp is inside an allowed freshness window;
  \item nonce has not been used before;
  \item body hash matches the received body;
  \item request signature verifies over the canonical envelope.
\end{itemize}

If these checks pass, OAN hands control to the application protocol. That
protocol may be A2A, MCP-adjacent tool invocation, a domain API, or a custom
workflow.

\section{Security Properties}

OAN targets the following properties:

\begin{itemize}[leftmargin=*]
  \item \textbf{Provenance}: accepted identity representations trace to a
  Root-authorized acceptance event.
  \item \textbf{Integrity}: tampered identity representations fail hash or proof
  verification.
  \item \textbf{Freshness}: stale, revoked, or superseded versions are not
  admissible after synchronization.
  \item \textbf{Authorization soundness}: compliant Discovery nodes do not
  index packages outside their authorized domains.
  \item \textbf{Replay resistance}: signed upstream writes and invocation
  requests bind timestamp, nonce, target, path, and body hash.
  \item \textbf{Controller non-impersonation}: a resource controller field is
  not sufficient by itself; externally controlled registration requires a
  controller authorization proof over the exact registration context.
  \item \textbf{Application-authentication separation}: login-style challenge
  verification uses Root-accepted public-key evidence, but it does not authorize
  resource registration or replace VC and package verification.
  \item \textbf{Role accountability}: each privileged decision is attributable
  to an authorized infrastructure identity.
\end{itemize}

\section{Threat Model}

\subsection{Malicious or Unauthorized Registrar}

An unauthorized Registrar may attempt to submit an identity to Root. Root
rejects it because the signed upstream envelope signer is not active in Root
authorization state. A malicious but authorized Registrar may issue poor
credentials; OAN can detect invalid signatures, wrong subjects, and policy
violations, but semantic review quality remains a governance problem.

\subsection{Compromised Root or Governance}

A compromised Root signing boundary can issue apparently valid acceptance
records, packages, or infrastructure credentials, while governance capture can
authorize malicious infrastructure. Ordinary in-domain signature verification
cannot detect these events because the affected authority is itself a trust
anchor. Threshold governance, protected Root keys, audit logs, revocation, and
recovery procedures reduce exposure; cross-domain validation requires a future
federation profile.

\subsection{Tampered Package Distribution}

An attacker controlling a CDN path may return modified documents or metadata.
Discovery recomputes hashes and verifies Root proof. Tampering should cause
package rejection.

\subsection{Overbroad Discovery Exposure}

A Discovery node may try to expose packages outside its authorized domains.
Compliant Discovery software should filter such packages before indexing.
Relying parties can also verify Discovery authorization state. Non-compliant
Discovery behavior remains auditable if returned candidates include provenance
that contradicts authorization policy.

\subsection{Stale Lifecycle Cache}

A Registrar, Discovery, or VC issuer node may miss bulletin events because of
network failure or local indexer downtime. The service should expose sync lag,
last processed event sequence, and cache health. Authorization-sensitive
actions should fail closed when lifecycle state is unavailable or too stale.

\subsection{Replay and Wrong-Target Invocation}

An attacker may replay a previously signed request or redirect a signed request
to another Service Agent. Timestamp windows, nonce stores, target DID binding,
path binding, and body hash binding reduce these risks.

\subsection{Controller Impersonation}

An attacker may publish a resource DID Document whose \texttt{controller} field
names a well-known controller DID. OAN does not treat that string as authority.
For externally controlled resources, the Registrar and Root must verify a
controller authorization proof signed by the controller over the resource DID,
controller DID, Registrar DID, document hashes, purpose, validity window, and
nonce. Without that proof, the claimed controller relationship is only
untrusted metadata.

\subsection{Login-Key Substitution}

An attacker may try to log in to an application by uploading a DID Document
that binds a chosen DID to the attacker's own public key. Applications that use
OAN Identity for login should therefore obtain accepted public-key evidence
from Root's identity directory or an equivalent Root-accepted projection and
then verify a fresh challenge signature. A user-uploaded DID Document alone is
not a sufficient login trust anchor.

\subsection{Semantic Misrepresentation}

An Agent may claim a capability that it does not perform well. OAN can bind and
govern the identity claim, but it cannot fully prove business quality or truth
of all semantic claims. Reputation, testing, certification, and runtime policy
must complement OAN.

\section{Compatibility and Adjacent System Boundaries}

OAN supplies identity and discovery evidence to MCP, A2A, ANP, and other Agent
protocols rather than replacing their task and message
semantics~\cite{mcp_spec,a2a_spec,did_wba_spec}.

MCP can use OAN to verify the identity and authorization context of tool
servers or clients before tool access. An MCP client may check that a server
endpoint belongs to a Root-accepted resource identity. An MCP server may use OAN
credentials to decide whether a caller is allowed to access a tool.

A2A can use OAN to verify Agent identity and discovery provenance before task
exchange. An A2A Agent Card or capability description can be mapped into the
OAN DID-style identity representation, while A2A remains responsible for task
state and message semantics.

ANP-like routes and Agent DID methods can be mapped to OAN identity documents
and lifecycle checks. OAN's distinctive contribution is not only having an
Agent identifier; it is the governed identity lifecycle, Discovery
authorization, package verification, and pre-connection trust guard around that
identifier.

\subsection{Registry, Directory, and Schema Boundary}

Several adjacent systems solve neighboring but different problems:

\begin{itemize}[leftmargin=*]
  \item MCP registries make MCP servers and server metadata discoverable, but
  they do not by themselves define DID/VC-based governance, Root acceptance, or
  infrastructure authorization~\cite{mcp_registry}.
  \item Skill marketplaces package and distribute Skills, while OAN gives a
  Skill a governed resource identity, lifecycle state, and verifiable discovery
  path.
  \item Agent schema frameworks provide vocabulary for Agentic records,
  capabilities, Skills, and relationships~\cite{agntcy_oasf}. OAN can embed or
  reference such records as package metadata.
  \item Agent Directory Service (ADS)-style systems emphasize distributed Agent directory and discovery
  functions~\cite{agntcy_ads_overview}. OAN emphasizes the trust path before and
  around discovery: Registrar onboarding, Root acceptance, package proof,
  governance-bounded Discovery, and signed verification.
  \item Agent Name Service (ANS)-style systems emphasize names for
  Agents~\cite{huang2025ans}. OAN
  can carry names as aliases, but its core identifier is the \texttt{did:oan}
  resource DID.
\end{itemize}

These systems address complementary concerns: external schemas describe
resources, ADS supports directory functions, ANS supports naming, MCP defines
tool interaction, and Agent runtimes manage execution. OAN adds the admission,
provenance, lifecycle, and verification context around resource publication,
discovery, and interaction.

\subsection{MCP Mapping}

An MCP server can be represented as an OAN MCP Server resource or as a service
endpoint inside a \texttt{did:oan} identity representation. MCP tool metadata can be mapped
to OAN service metadata and capability tags. Before MCP initialization or tool
execution, a client can query OAN Discovery, verify the Discovery response,
verify the Root package, and then bind the MCP endpoint to the verified
resource identity. MCP remains responsible for tool listing, tool invocation, resource
access, and streaming behavior.

\subsection{A2A Mapping}

An A2A Agent Card can be mapped to an OAN identity representation by binding
Agent identifier, service endpoints, capability declarations, and verification
methods. OAN Discovery can serve as the pre-session discovery layer. After the
candidate is verified, A2A can take over task negotiation, message flow, task
state, and long-running collaboration. This avoids forcing A2A to solve Root
governance and Discovery authorization internally.

\subsection{ANP and DID Method Mapping}

ANP-like route structures and Agent DID methods can be used as identifier and
networking substrates. OAN can consume such identifiers if they can be resolved
into verification methods and service metadata. The OAN-specific part is the
registration path, Root acceptance, verified package, and authorized Discovery
semantics around that identifier. Table~\ref{tab:integration-boundaries}
summarizes the resulting division of responsibility.

\begin{table}[t]
\centering
\caption{Protocol Integration Boundaries}
\label{tab:integration-boundaries}
\begin{tabular}{@{}lll@{}}
\toprule
Protocol & Keeps responsibility for & OAN provides \\
\midrule
MCP & tools and resources & peer identity trust \\
A2A & task exchange & trusted discovery \\
ANP & Agent networking & lifecycle governance \\
DID/VC & primitives & Agent-specific policy \\
Domain API & business action & pre-connection guard \\
\bottomrule
\end{tabular}
\end{table}

\section{Integration Depth and Adoption Interfaces}

OAN defines integration depth as the extent to which a native Agent protocol
uses governed OAN identity material in its lifecycle and message path. The
depth is independent of whether the integration is embedded in an Agent,
provided by a local tool, or enforced at a network boundary. The same resource
DID and public evidence model can therefore support several adoption paths
without introducing a second identity authority. Each depth is a conformance
claim for a named protocol and deployment profile: it must identify the objects,
message paths, key boundary, and fail-closed checks that are actually enforced.
A Runtime, Sidecar, Gateway, or Skill is an adoption interface, not evidence
that the surrounding application has automatically reached a particular depth.

\subsection{L0: Publication and Discovery}

L0 covers registration and discovery. The resource has a valid OAN identity
representation, a governed registration record, a discoverable endpoint or
artifact reference, and the public materials needed to resolve its identity.
A consumer can associate
the selected MCP Server or A2A Agent with an OAN DID and inspect its published
metadata. L0 establishes an identity and discovery binding, but it does not
make an arbitrary later protocol message an authenticated OAN message.

\subsection{L1: Evidence-Aware Interaction}

L1 adds verification before protocol interaction. The consumer verifies the
DID Document, registration credential, Root-verified package, Discovery
provenance, current version, and endpoint binding under an explicit trust
policy. The native protocol may continue to use its ordinary transport and
authorization mechanisms. L1 is satisfied when the OAN evidence is checked and
the result is applied to the interaction decision; a transport channel or a
plain identifier string alone is not an L1 OAN identity proof.

\subsection{L2: Mutual Identity and Message Protection}

L2 applies OAN identity to a real two-sided protocol path. Each resource proves
control of its own resource key, and the session binds the expected peers,
purpose, audience, freshness, and replay state. The request is verified before
the native handler executes, and the response, result, stream event, callback,
or artifact is bound to the request and the correct responding resource before
it is accepted as trusted output. Authorization is evaluated at the same
protected boundary. A controller, proxy, or channel identity cannot substitute
for the resource identity that signs the business message.

For MCP, an L2 profile binds the server and client resources to tool selection,
arguments, result or error, and stream or cancellation state. For A2A, it binds
the Agent Card and task participants to task identifiers, artifacts, event
ordering, callbacks, and terminal state. The native protocols retain their
wire and business semantics; OAN supplies the identity, evidence, and message
binding rules around them. An implementation should therefore publish its
supported MCP or A2A profile explicitly, including which synchronous, streaming,
callback, cancellation, resource, task, or artifact paths are covered;
unsupported paths must not be presented as L2.

\subsection{L3: Extended Operational and Business Controls}

L3 extends the L2 foundation with controls required by a larger deployment or a
more sensitive business process. These controls may include bounded delegation,
tenant and organizational policy, protected private-network ingress, remote or
hardware-backed key providers, confidential payload handling, stronger audit
requirements, and runtime or instance assurances. Such controls add policy and
deployment evidence around OAN identity; they do not change the resource-level
meaning of a DID, controller relationship, or business signature.

\subsection{Adoption Interfaces}

\begin{figure*}[!ht]
\centering
\includegraphics[width=0.8\linewidth]{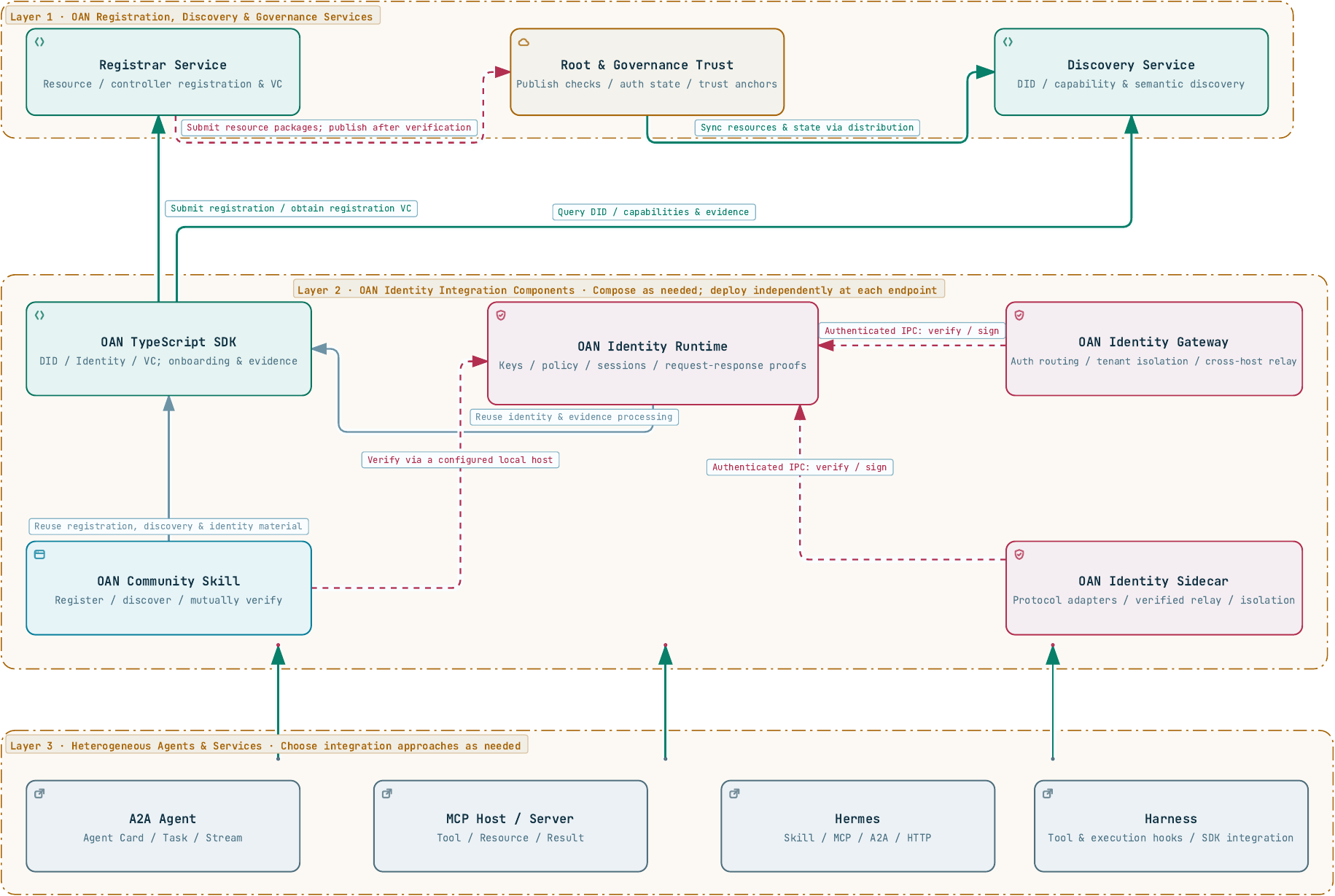}
\caption{Identity adoption interfaces and their dependencies. Registration and
Discovery use shared SDK facilities; Gateway and Sidecar request controlled
verification and signing from Runtime. Endpoint integrations compose these
capabilities without replacing native protocol semantics.}
\label{fig:yp-identity-integration}
\end{figure*}

Figure~\ref{fig:yp-identity-integration} locates the adoption interfaces between
OAN infrastructure and heterogeneous protocol endpoints. The SDK supplies
shared identity and evidence functions; Skill orchestrates workflows, Runtime
controls identity use, and Gateway and Sidecar enforce protected communication
boundaries through Runtime interfaces.

The Skill interface is an orchestration surface for assistants that can execute
local skills or tools. It handles user intent, registration and Discovery
workflows, identity-material handoff, and public challenge exchange by calling
the local identity facilities. It may also request a protected call through a
declared profile, but Skill execution alone cannot force unrelated business
traffic through an OAN boundary.

The Identity Runtime is the edge execution boundary for identity use. It owns
identity selection, controlled key-provider calls, evidence verification,
policy decisions, session state, replay state, and audit inputs. A Runtime is a
consumer of OAN trust evidence, not a replacement for Root, Registrar, or
Discovery, and it must not expose an unrestricted private-key signing primitive
to an Agent or model.

A Sidecar provides a low-intrusion enforcement boundary around an existing MCP
or A2A service. It verifies ingress, obtains controlled signing or verification
decisions from the Runtime, protects egress, and supplies only verified caller
context to the backend. Backend credentials and private keys remain outside
the business process, and direct unauthenticated bypass of the protected
endpoint is rejected by deployment policy.

A Gateway provides a forwarding and organizational boundary for private
networks, multi-tenant deployments, and delegated access. Its protected
envelope distinguishes the original caller, the proxy, and the target
resource, and it binds delegation and payload digests to the forwarded
operation. Gateway channel authentication is therefore separate from the OAN
business identity carried by the resource message.

These interfaces are composable. A Skill can call a Runtime; an MCP or A2A
implementation can embed the Runtime; a legacy service can be protected by a
Sidecar; and a Gateway can connect tenants or private domains while preserving
the resource identity chain. Integration depth is assessed for each declared
protocol path and deployment profile according to its identity bindings and
enforcement behavior. A client library or deployment component alone does not
establish that depth.

\section{Implementation Profiles}

OAN implementations do not need to implement every role at once.
Table~\ref{tab:implementation-profiles} defines the expected depth of each
implementation profile.

\begin{table}[t]
\centering
\caption{Implementation Profiles}
\label{tab:implementation-profiles}
\begin{tabular}{@{}lll@{}}
\toprule
Profile & Required capabilities & Typical implementer \\
\midrule
Agent & verify Discovery, sign calls & Agent runtime \\
Discovery & sync packages, sign query & directory operator \\
Registrar & draft, credential, submit & onboarding operator \\
Root & enforce, accept, sign package & trust-domain operator \\
Adapter & map OAN to MCP/A2A/ANP & protocol integrator \\
Full stack & all infrastructure roles & trial network \\
\bottomrule
\end{tabular}
\end{table}

The profiles allow staged adoption. A partner can begin with an Agent
profile, later operate a Discovery node, and only eventually operate Registrar
or Root infrastructure. This makes OAN deployable as a layered ecosystem rather
than an all-or-nothing platform.

\section{Implementation Boundary}

The reference implementation separates responsibilities along the same
boundaries as the protocol model. Root implements acceptance, proof issuance,
package publication, distribution coordination, and trust-domain services.
Registrar implements governed onboarding, DID Document validation, registration
review, and Root submission. Discovery implements package synchronization,
structured indexing, derived semantic indexing, scoped query, and signed
Discovery results. The Trust Indexer converts low-frequency governance
events into runtime-readable node authorization state. Client libraries help
applications generate OAN Identity material, create signed envelopes, verify
DID Documents and credentials, and consume Discovery responses without
reimplementing every cryptographic rule.

Repository layout, build scripts, database schemas, and operator runbooks are
not part of the protocol. The named components are reference realizations of
Root, Registrar, Discovery, governance indexing, and client-side verification
roles. Identity Runtime, Sidecar, Gateway, and Skill components consume OAN
evidence and policy at deployment boundaries; they do not acquire the
authority of infrastructure roles.

\subsection{Normative Boundary}

The technical design distinguishes normative protocol requirements from
reference implementation choices. Normative requirements include signature
coverage, hash binding, nonce and timestamp validation, Root proof
verification, current-version checks, Discovery authorization filtering,
failure reason reporting, and fail-closed behavior for security-critical
verification. An implementation that omits these checks is not implementing the
OAN trust model, even if it exposes similar HTTP routes.

Reference implementation choices include concrete database engines, queue
implementations, local cache layouts, index structures, deployment automation,
observability surfaces, and default retry policies. These choices may vary across
operators as long as the externally visible verification semantics and
conformance results remain consistent.

Human-facing portals, maps, dashboards, and documentation sites are
non-normative presentation surfaces. The normative protocol semantics are DID
Document validation, resource-type consistency, Root package verification,
authorization-domain enforcement, Discovery response signing,
lifecycle-state handling, and verification failure behavior.

\section{Persistence and Operational State}

Root, Registrar, Discovery, and CDN each maintain state with different
criticality.

\begin{itemize}[leftmargin=*]
  \item Root state is authoritative for authorization, accepted versions,
  publication, and bulletin facts.
  \item Registrar state is authoritative for draft workflow, subject-control
  evidence, credential issuance, and submission attempts.
  \item Discovery state is authoritative for local synchronization cursor,
  package verification results, rejected package reasons, and query indexes.
  \item CDN state is authoritative for distributed package availability and
  manifest history, but not for trust decisions.
\end{itemize}

Implementation backends are non-normative. A reference implementation may use
different local or production storage engines, but a conforming deployment must
preserve the state needed to recover authoritative and derived trust decisions.

Operational state is classified by authority and recoverability. Accepted Root
records, resource versions, Root proofs, and Registrar submissions require
durable preservation. Discovery synchronization cursors, structured indexes,
and semantic indexing status require either durable storage or a recovery path
from accepted evidence. Rejected or skipped package records support recovery
and diagnosis. Operational statistics and public metrics are derived views;
their retention policies do not define protocol acceptance. Disposable response
caches, UI rendering state, and health probes can be recomputed. Replay
protection state must remain effective throughout the applicable request-validity
window, including across restart.

A conforming deployment preserves enough authoritative state and recovery
information to reconstruct Root acceptance, Registrar submission history,
Discovery visibility, and semantic indexing progress after restart. Storage
products and optional public reporting remain deployment choices.

\section{Performance Envelope}

The performance measurements exercise the asynchronous
registration-to-discovery path under four combinations of resource composition
and logical topology. Single-node workloads use one Registrar and one Discovery
node; logical multi-node workloads use three Registrars and two Discovery
nodes. Both include one Root, one CDN node, one publication worker,
PostgreSQL~17, and NATS JetStream. Agent-only and mixed-resource workloads use
the same verification and publication path.

The services ran as release builds on one Windows host over loopback. Each
scenario and scale started from fresh isolated state. Four client processes
maintained 48 total in-flight registrations; DID verification used four workers
and a queue capacity of eight, and the relevant PostgreSQL pools were set to
eight connections. Resource construction, compilation, and service startup
preceded the common measurement start. The observer sampled deduplicated Root,
CDN, and per-Discovery counts once per second without issuing per-resource
queries in the timed interval.

Each of the four workloads ran five times at three target scales, for 60
isolated measurements. The logical multi-node Agent workload uses equal
per-Registrar counts, yielding actual totals of 498, 1998, and 4998; the other
workloads use 500, 2000, and 5000 resources. These experiments characterize
logical role composition and structured lifecycle completion on one host.
Physical scale-out, production capacity, Agent task execution, and
semantic-index convergence are outside the measured scope.

\begin{figure}[!ht]
\centering
\includegraphics[width=0.47\textwidth]{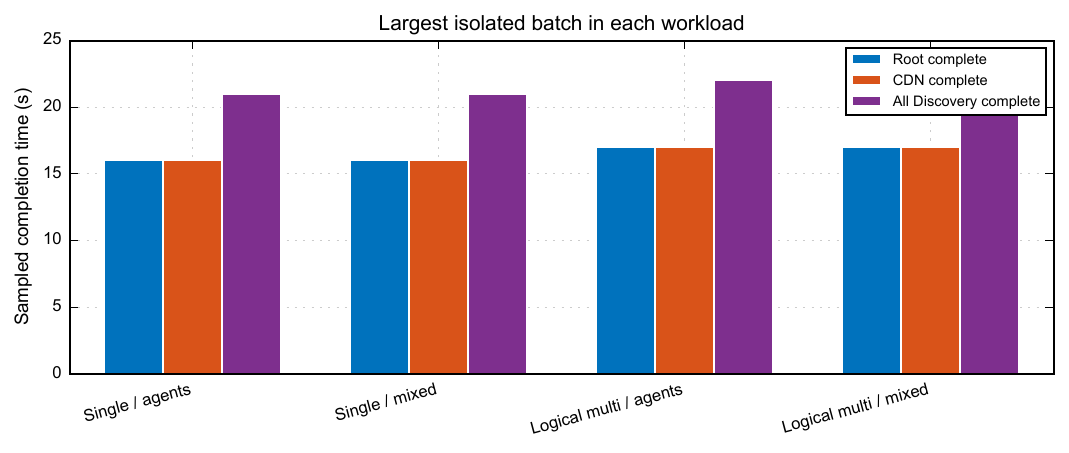}
\caption{Median Root, CDN, and all-Discovery completion times at the largest
scale. Ranges cover five isolated runs; timestamps come from one-second
sampling.}
\label{fig:yp-pipeline-completion}
\end{figure}

\paragraph{Timing contract}
For a batch of $N$ resources, let $T_{\mathrm{submit}}$ denote the time until
all client requests have returned successfully. Let $T_{\mathrm{root}}$,
$T_{\mathrm{cdn}}$, and $T_{d_i}$ denote the first samples at which Root, CDN,
and Discovery target $i$ reach $N$ new resources. The completion contract is
\begin{align*}
 T_{\mathrm{discovery}} &= \max_i T_{d_i},\\
 T_{\mathrm{e2e}} &= \max(T_{\mathrm{root}},T_{\mathrm{cdn}},
                          T_{\mathrm{discovery}}),\\
 R_{\mathrm{root}} &= N/T_{\mathrm{root}},\qquad
 R_{\mathrm{e2e}} = N/T_{\mathrm{e2e}},
\end{align*}
with times expressed in seconds from the common start. The one-second observer
makes the three completion times sampled upper bounds with approximately one
sampling interval of uncertainty. Publication and Discovery synchronization
overlap client submission, so differences between these times describe
pipeline progression rather than additive service times.
Figure~\ref{fig:yp-pipeline-completion} shows the median stage-completion times
and observed ranges at the largest scale.

\begin{table}[!ht]
\centering
\caption{Largest-Scale Structured Pipeline Results}
\label{tab:yp-pipeline-largest}
\begin{tabular}{@{}lrrrr@{}}
\toprule
Workload & $N$ & $T_{\mathrm{e2e}}$ (s) & $R_{\mathrm{root}}$ & $R_{\mathrm{e2e}}$ \\
\midrule
Single / agents & 5000 & 21.021 & 311.51 & 237.86 \\
Single / mixed & 5000 & 21.011 & 311.58 & 237.97 \\
Logical multi / agents & 4998 & 22.021 & 293.40 & 226.97 \\
Logical multi / mixed & 5000 & 24.077 & 293.98 & 207.67 \\
\bottomrule
\end{tabular}
\end{table}

\begin{figure}[!ht]
\centering
\includegraphics[width=0.47\textwidth]{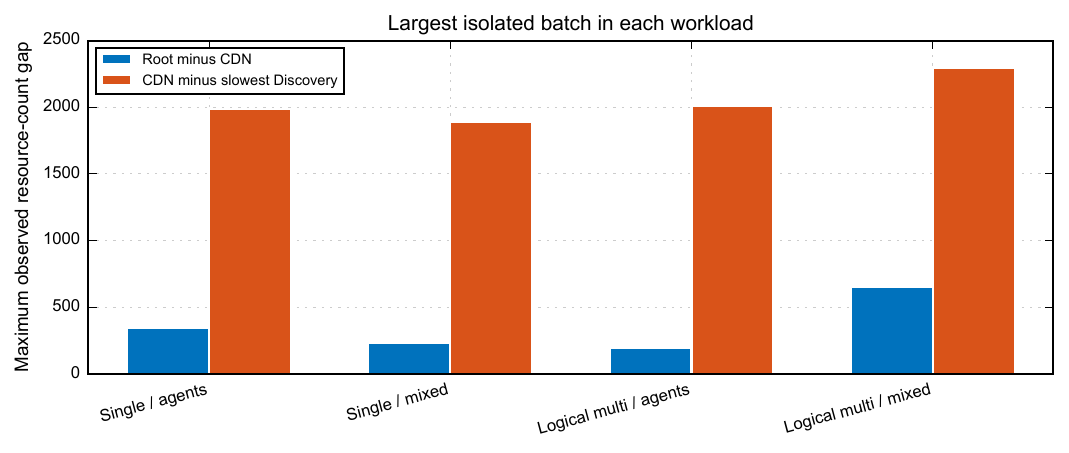}
\caption{Median maximum sampled count gaps between Root and CDN, and between
CDN and the slowest Discovery target, at the largest scale. Error bars show
the observed range across five runs.}
\label{fig:yp-pipeline-backlog}
\end{figure}

\paragraph{Backlog and completeness}
All 60 runs reached the expected deduplicated counts at Root, CDN, and every
Discovery target. Median sampling-error counts were zero in all 12
scenario--scale groups. At the largest scale, median maximum Root--CDN count
gaps are 345, 230, 191, and 650 resources for the four workloads in
Table~\ref{tab:yp-pipeline-largest}; median CDN--slowest Discovery gaps are
1984, 1889, 2013, and 2296. Figure~\ref{fig:yp-pipeline-backlog} shows these
sampled maxima. They locate transient accumulation between stages but are not
exact queue depths or per-event residence times.

The results distinguish client completion, Root acceptance, durable
publication, and Discovery convergence while preserving a single correctness
condition: every target must reach the expected resource set. They establish a
structured lifecycle baseline under the recorded configuration. The real-model
semantic correctness gate could not run because the local PostgreSQL instance
lacked pgvector; no semantic quality or convergence result is inferred. A
strict optimization A/B result is also omitted because the earlier binaries
were not rerun under this measurement contract.

Five independent security and lifecycle gates passed on fresh topologies:
end-to-end trusted invocation, trusted negative cases, authorized-domain
isolation, controller continuity, and the four-resource matrix. These gates
cover invariants I1--I7, including stale-controller rejection, nonce replay,
target and body-hash binding, out-of-domain non-exposure, and compatibility of
controller transitions with CDN and Discovery. The earlier degraded-baseline
ablation below provides complementary mechanism-isolation evidence.

\begin{figure}[!ht]
\centering
\includegraphics[width=0.47\textwidth]{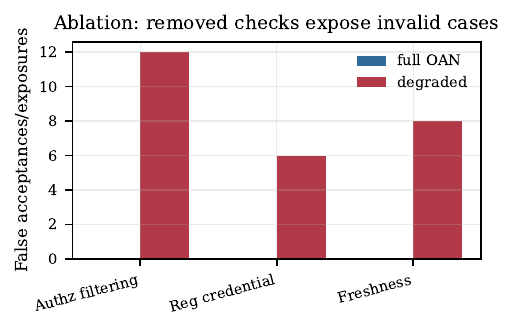}
\caption{Real-system degraded-baseline ablation. Removing authorization
filtering, registration-credential verification, or freshness checks causes
false exposure or false acceptance in the degraded variants, while the full OAN
path rejects all tested invalid cases.}
\label{fig:yp-ablation}
\end{figure}

The ablation results in Figure~\ref{fig:yp-ablation} were obtained on copied
reference-service workspaces so that the normal implementation remained
unchanged. Each degraded variant
removed one mechanism and then exercised the corresponding Root-node or
Discovery-node code path over the same generated cases. The full path produced
zero false acceptances or false exposures across the tested invalid cases,
whereas the degraded variants exposed all cases designed to target the removed
check. These results support the claim that OAN's authorization, credential,
and freshness checks are functional trust mechanisms rather than merely
descriptive metadata.

\subsection{Core-Service Resource Footprint}
\label{sec:yp-resource-footprint}

Resource behavior under load was measured in a separate 5000-resource
single-Registrar run so that process sampling did not perturb the formal
throughput matrix. The isolated topology remained running for 600~s after
structured completion. Table~\ref{tab:yp-dynamic-footprint} reports working
sets at the last observation and the peak equivalent CPU demand of each
managed role. PostgreSQL is shown with a host-wide process aggregate and is
therefore not directly comparable with one managed node process.

\begin{table}[!ht]
\centering
\caption{Dynamic Footprint in the 5000-Resource Run}
\label{tab:yp-dynamic-footprint}
\footnotesize
\begin{tabular}{@{}lrr@{}}
\toprule
Role & Last working set (MiB) & Peak CPU cores \\
\midrule
Root & 41.5 & 2.66 \\
Registrar & 33.9 & 1.37 \\
Discovery & 76.8 & 0.62 \\
CDN node & 18.8 & 0.39 \\
CDN publisher & 16.9 & 0.18 \\
NATS & 29.6 & 0.11 \\
PostgreSQL host aggregate & 371.7 & n/a \\
\bottomrule
\end{tabular}
\end{table}

\begin{figure}[!ht]
\centering
\includegraphics[width=0.47\textwidth]{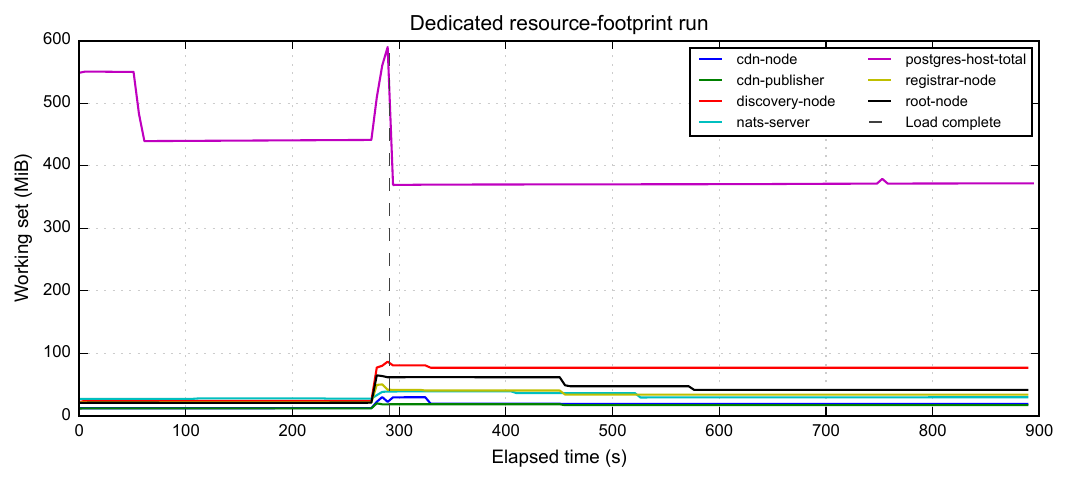}
\caption{Working-set observations for the dedicated 5000-resource footprint
run, including the post-completion settling window. PostgreSQL uses a
host-wide aggregate and a different accounting scope.}
\label{fig:yp-resource-footprint}
\end{figure}

Root peaked at 64.7~MiB and settled to 41.5~MiB at the ten-minute observation;
the other managed roles likewise showed no continuing post-load growth. This
result addresses the earlier Root memory-amplification concern for the tested
configuration, but it is not a general leak-freedom proof. Figure~\ref{fig:yp-resource-footprint}
shows the recorded working-set series.

The following populated Linux snapshot addresses a different question: the
steady-state composition of a production-like deployment with semantic
inference enabled and more than 22,000 registered resources. It should not be
combined numerically with the isolated dynamic run.

OAN separates lightweight identity and publication services from shared
storage and semantic inference. A steady-state Linux deployment containing two
Registrar services, two Discovery services, and the Root, Trust Indexer,
CDN, publication, and messaging services maintained more than 22,000
registered resources. Table~\ref{tab:yp-resource-memory} reports a
resident-memory snapshot at this population. Registrar and Discovery values
are the arithmetic means of their two service instances. The measurements
describe a populated, low-traffic operating state, separately from the
batch-throughput workloads above.

\begin{table}[!ht]
\centering
\caption{Core-Service Memory at More Than 22,000 Registered Resources}
\label{tab:yp-resource-memory}
\begin{tabular}{@{}lr@{}}
\toprule
Component & Memory (MiB) \\
\midrule
Root & 41.5 \\
Registrar (mean per instance) & 22.6 \\
Discovery (mean per instance) & 104.4 \\
Trust Indexer & 8.2 \\
CDN Node & 9.5 \\
CDN Publisher & 9.7 \\
NATS & 12.4 \\
Discovery Embedding & 1320.4 \\
PostgreSQL (service cgroup) & approximately 1170 \\
\bottomrule
\end{tabular}

\smallskip
\parbox{\columnwidth}{\footnotesize Process entries use resident set size;
PostgreSQL uses service-cgroup memory, including child processes and charged
page cache. These accounting scopes differ and should not be summed as
exclusive physical memory. One MiB is $2^{20}$ bytes.}
\end{table}

\paragraph{Model baseline and resource-dependent storage.}
The semantic service's approximately 1.29~GiB resident footprint is dominated
by the loaded embedding model, inference runtime, and reusable working
buffers. This is primarily a shared inference cost, not a resident model or
vector collection allocated for each registered resource. Resource vectors
are persisted by Discovery in PostgreSQL; inference memory depends on model
choice and active batch/concurrency settings rather than requiring a copy of
the full resource population. For scale intuition, one 768-dimensional
float32 vector contains 3072 bytes: 22,000 such vectors contain approximately
64.5~MiB of numerical payload per indexed copy, excluding record and index
overheads. This is a storage calculation, not a measured RAM increment.

PostgreSQL's approximately 1.14~GiB cgroup footprint likewise must not be
interpreted as its empty-database baseline. The inspected configuration uses
128~MiB of shared buffers. Cgroup accounting showed approximately 40~MiB of
anonymous memory, 137~MiB of shared-memory pages, and 968~MiB of other
file-backed pages; the last category includes reclaimable database-file cache.
Shared-memory pages are included in the reported file total and are not
counted twice here. The baseline consists of database processes, connection
workspaces, and configured shared buffers; the resource-dependent component
includes cached resource records, credentials, vectors, and index pages.
An empty-population control was not measured, so the snapshot does not assign
an exact incremental RAM cost to the 22,000-plus resources. Persistent corpus
growth is distinct from a requirement to keep every record resident, allowing
the storage working set to be managed independently of the small application
processes.

\paragraph{Steady-state CPU demand and scaling implications.}
After a five-minute settling interval, a continuous 180-second observation
of this deployment measured an average host demand of approximately
0.128 CPU cores, equivalent to 3.2\% of a four-core machine. This host-level
envelope includes the two Registrars, two Discovery services, Root, and
supporting services; PostgreSQL accounts for most of the measured demand.
Equivalent cores are accumulated CPU seconds divided by elapsed wall-clock
seconds. The observation characterizes background operation, not peak
registration or inference demand.

The small application footprints and low steady-state CPU demand leave
compute headroom for active registration, discovery, and identity verification.
Periodic statistics caches coalesce repeated scans, while bounded retrieval
and durable storage avoid requiring application processes to retain the entire
resource corpus. Shared semantic inference amortizes model residency across
resources, and storage, inference, and protocol services can be provisioned
independently. These properties provide an economical basis for expanding
Agent populations; throughput and latency under increased activity remain
the complementary dimensions of capacity planning.

\subsection{Scenario-Based Identity and Interoperability Validation}

Three controlled application cases exercise the adoption interfaces on native
A2A and MCP paths: domestic e-commerce, cross-domain data verification, and
AI-for-Science collaboration. The cases contain six, six, and eight actors,
with six, six, and nine declared business interaction edges respectively. Each
actor is associated with an OAN resource identity and registration evidence;
the interaction graph associates each operation with its caller, target,
protocol, and expected responding resource. The application workloads make
these bindings observable across task and tool boundaries.

\paragraph{Identity and protocol binding.}
The case definitions compose Skill/SDK onboarding and discovery with
Runtime-owned identity selection, key use, session checks, and signing. Gateway
and Sidecar boundaries are designed to enforce configured routes and obtain
cryptographic decisions through authenticated Runtime interfaces. A2A carries
task and result semantics, while MCP carries tool operations and native
results; application handlers receive verified caller context, not private
keys. The current evidence package passes 21 offline regression checks for
builds, types, protocol versions, scenario declarations, and P7--P9 aggregation
rules. Live per-edge execution has not been claimed because the three private
deployment manifests were not available in this run.

The intended multi-hop rule is checked by the offline fixtures: a service uses
its own resource identity when invoking a downstream tool, verifies the tool's
response, and binds its upstream response to the corresponding request.
Correlation identifiers do not substitute an upstream caller or controller
key for the downstream resource identity. Demonstrating this rule against the
three deployed applications remains part of the live evidence plan.

\paragraph{Streaming, cancellation, and recovery.}
The maintained scenario contracts define one A2A streaming edge, one remote
MCP cancellation edge, direct-backend bypass rejection, and Gateway/Sidecar
restart recovery for each case. The offline checks validate that these paths
and expected assertions remain present and version-compatible; they do not
show that remote streaming, cancellation, or restart recovery succeeded in a
live deployment. Those outcomes require the complete runtime evidence matrix.

\paragraph{Application constraints and interpretation.}
The e-commerce workload checks budget and merchant registration evidence;
the cross-domain workload checks source-record boundaries, digest agreement,
and the expected data domain; the scientific workload checks units, versions,
and result-summary consistency. Their negative cases test the corresponding
application decisions alongside the shared identity boundary. OAN registration
VCs supply the registration evidence, and the application payloads provide
controlled inputs for observing acceptance and rejection.

Together, the case artifacts specify how publication and discovery evidence is
carried into heterogeneous protocol paths. The offline result supports
regression confidence in the L0/L1 interfaces and the selected L2 contracts;
it does not establish live interoperability. Actor and edge counts describe
planned verification coverage, not throughput, concurrency limits, or
scientific-model accuracy, and the application checks do not constitute
domain-level validity assessments.

\section{Conformance Expectations}

Independent implementations should be able to pass conformance tests for:

\begin{itemize}[leftmargin=*]
  \item \texttt{did:oan} identifier parsing, routing-code checks, DID Document
  profile validation, and canonical hashing;
  \item DID Document and VC proof verification using the current Data Integrity
  proof profile;
  \item OAN Identity generation and import checks that keep private key
  material local while binding it to the public DID Document;
  \item controller authorization proof verification for externally controlled
  resources and self-controller handling for self-controlled resources;
  \item external identifier parsing and validation without treating native
  external identifiers as OAN DIDs;
  \item product-type metadata for Agent Service, Skill, MCP Server, and
  Tool/API resources;
  \item external artifact reference digest verification;
  \item registration credential proof verification;
  \item subject-control challenge verification;
  \item signed upstream envelope verification;
  \item Root package verification;
  \item Discovery response signature verification;
  \item authorized-domain validation and authorization;
  \item nonce and timestamp replay rejection;
  \item current-version and revocation handling.
\end{itemize}

Conformance tests should include rejection of malformed credentials, stale
packages, unauthorized Discovery exposure, and replayed requests. Successful
execution and deterministic rejection together provide evidence of conformance
with the OAN trust model.

Conformance should also distinguish product-specific expectations from common
trust expectations. A Skill conformance vector may not require a network
endpoint, but it must require artifact-reference integrity and semantic
metadata. An MCP Server vector should require endpoint binding and protocol
hints, but it should not require OAN to validate the full MCP tool list. A
Tool/API vector should require endpoint or schema-reference binding, while API
business semantics remain outside the OAN core.

\section{API Semantics}

OAN APIs expose semantic contracts for registration, acceptance, publication,
and Discovery. These contracts preserve verification behavior across
implementations while allowing deployment-specific interface choices.

OAN separates protocol evolution from API path naming. Version information in
protocol objects allows verification rules and data profiles to evolve without
requiring every change to introduce a new route prefix. Compatibility depends
on the supported object profiles and their verification semantics, rather than
on the presence of a version number in a URL.

The reference route profile groups interfaces by role:

\begin{itemize}[leftmargin=*]
  \item Root: \texttt{/root/...}
  \item Registrar: \texttt{/registrar/...}, \texttt{/resources/...},
  \texttt{/capability-tree}
  \item Discovery: \texttt{/discovery/...}, especially
  \texttt{/discovery/resources/...}
  \item CDN: \texttt{/cdn/resources/...}, \texttt{/cdn/documents/...},
  \texttt{/cdn/metadata/...}
\end{itemize}

Where incompatible interface changes require an explicit API version,
deployments can use content negotiation or versioned prefixes while preserving
the identity and evidence bindings defined by the applicable protocol profile.

\subsection{Root APIs}

Root APIs can be divided into management, verification, publication, and read
interfaces. Management interfaces update infrastructure authorization state,
authorized-domain state, and operational configuration. Verification
interfaces accept signed Registrar submissions and return deterministic accept
or reject results. Publication interfaces create or expose Root-verified
packages and publication cursors. Read interfaces expose status, bulletin
facts, accepted versions, and authorized infrastructure state.

Root write APIs should require signed upstream envelopes. Root read APIs may be
public or restricted depending on deployment, but the returned trust facts
should be verifiable or derived from verifiable Root state.
Root should not be treated as the public full-resource listing service. A Root
read interface may expose verifiable Root state, package material, cursors, or
accepted-version facts, but general resource enumeration and user-facing search
belong to Discovery. Public network statistics belong to a visibility profile,
not to Root protocol authority.

Root may also expose an identity-directory read interface that projects
Root-accepted controller or resource identities and verified public keys for
application authentication. This interface is a public-key discovery aid. It
does not authorize resource control by itself and does not replace DID Document
or VC verification.

\subsection{Registrar APIs}

Registrar APIs support draft creation, authorization-domain vocabulary
retrieval, capability-tag suggestion, authorized-domain selection,
subject-control challenge generation, controller-authorization challenge
generation when an external controller is used, proof submission, credential
issuance, and Root submission. A
Registrar API should preserve the relationship between a draft, its DID
document hash, its authorized domains, its subject-control proof, its
credential, and its Root submission result.

Registrar should also expose enough status for a resource provider to understand
why a draft cannot progress. For example, missing authorized domains,
out-of-scope domains, missing capability tags, failed DID control, expired
challenge, credential issuance failure, and Root rejection are different errors
and should not be collapsed into one generic failure.

\subsection{Discovery APIs}

Discovery APIs support status, synchronization, package inspection, rejected
package inspection, and query. The resource-oriented route profile uses
\texttt{/discovery/resources/...} and resource index/query routes to expose
heterogeneous resource types through a shared query model. Query responses
should be signed and should
include enough provenance for a User Agent to verify selected candidates.
Discovery should not expose packages that have not passed package verification
and authorization filtering.

If a Discovery node implements semantic matching, the API should make the
semantic profile observable without making it part of the trust proof. For
example, a response may include matched labels, embedding profile name,
semantic score, or explanation fields inspired by SLM-enhanced indexing methods
such as GRAIL~\cite{xu2026grail}. These fields should be signed as response
metadata for auditability, but verifiers must still base trust acceptance on
Root package evidence, Discovery authorization, and candidate provenance.

Human-facing Discovery inspection surfaces are non-normative views over this
API family. They may show resource DID, version, capability metadata,
description, and DID Document material, but they do not replace SDK or client
verification. Their protocol value is to keep Discovery output inspectable by a
relying party before invocation.

\subsection{CDN APIs}

CDN APIs expose manifests, packages, documents, metadata, and publication
history. CDN responses should be treated as distribution data, not trust
decisions. Consumers should validate Root proof and hashes after fetching CDN
content.

\section{Error Taxonomy}

Precise errors are part of interoperability. Table~\ref{tab:error-taxonomy}
summarizes common error categories.

\begin{table}[t]
\centering
\caption{Representative Error Categories}
\label{tab:error-taxonomy}
\scriptsize
\begin{tabular}{@{}p{0.2\columnwidth}p{0.34\columnwidth}p{0.34\columnwidth}@{}}
\toprule
Category & Example & Expected behavior \\
\midrule
syntax & malformed object & reject before trust checks \\
crypto & bad signature & reject and log signer \\
authz & inactive Registrar & reject as unauthorized \\
freshness & stale timestamp & reject as expired \\
replay & repeated nonce & reject and audit \\
binding & wrong subject DID & reject as mismatch \\
scope & out-of-domain submission or package & reject or do not index \\
state & superseded version & hide from query \\
publication & package not synchronized & return pending or no result \\
semantic & semantic index pending or failed & fall back or mark recoverable \\
metrics & stale public cache & show stale status without changing trust \\
portal & public display unavailable & keep protocol services independent \\
\bottomrule
\end{tabular}
\end{table}

Errors should be safe to expose. External responses may avoid leaking sensitive
operator details, but internal logs should preserve enough information for
audit and debugging.

A missing public search result should not be treated as one error class. It may
result from Registrar submission rejection, Root acceptance failure,
publication lag, Discovery out-of-scope filtering, package verification
failure, structured-index lag, semantic-index lag, query mismatch, or stale
portal cache. Implementations should expose enough status to separate these
cases without leaking private operational evidence.

\section{Consistency Model}

OAN does not require every component to become globally consistent
instantaneously. The architecture uses a staged consistency model.

\subsection{Root Consistency}

Root is authoritative for accepted resource identity versions, archived
representations, package generation, and Root-issued protocol credentials.
Infrastructure authorization is a joint state derived from active governance
status and valid Root-issued VC state. Within one Root deployment, state
transitions should be transactional enough to avoid accepting a version without
a corresponding archive and publication state. Root should also avoid
ambiguous current-version state for the same resource DID and should fail
closed when its governance-derived infrastructure cache is stale beyond the
configured trust window.

\subsection{Publication Consistency}

Publication from Root to CDN and Discovery can be asynchronous. A newly
accepted Agent may not be immediately queryable. This is acceptable if the
system exposes publication cursors, synchronization status, and retry behavior.
The important property is monotonic progress without silently losing accepted
packages.

\subsection{Discovery Consistency}

Discovery is eventually consistent with Root publication. It should never
compensate for staleness by accepting unverifiable packages. If it has not yet
synchronized a new accepted version, it may return no result or an older
current result depending on Root state and cursor timing, but it should not
return a package that fails current-version checks once it has learned the
superseding state.

\subsection{Discovery Backfill and Semantic Index Recovery}

A newly authorized Discovery node should be able to converge from an empty
local index to the set of Root-accepted resources within its authorized scope.
This backfill path should be idempotent: rerunning it must not duplicate
resources, widen authorization scope, or overwrite newer verified state with
older state. Backfill progress should be observable through Root accepted
counts, Discovery synchronized counts, structured-indexed counts,
semantic-indexed counts, pending or failed semantic-index records, and cursor
lag.

Structured synchronization and semantic indexing are allowed to be decoupled.
A semantic-index upsert failure should not roll back an already verified
package synchronization, but it must leave durable retry state. Otherwise, a
Discovery node may appear synchronized at the structured index level while
remaining permanently incomplete for intent-based search. The required safety
property is recoverable convergence: every accepted and in-scope package is
either indexed semantically, marked pending, marked failed with reason, or
excluded by an explicit authorization or lifecycle rule.

This decoupling is visible to operators but should not change the external trust
semantics of Discovery. DID lookup, structured filtering, package provenance,
and signed Discovery responses can be correct before semantic ranking has fully
caught up. Semantic indexing is a derived search capability over already
accepted structured state, not a precondition for Root acceptance or a
replacement for package verification.

Recoverable convergence is independent of the embedding model, vector
database, retry interval, or queue implementation. Implementation profiles
choose these mechanisms while preserving observable state and recovery
semantics.

\section{Key Management Considerations}

Every privileged role depends on signing keys. Key management is therefore a
first-class operational issue.

\subsection{Root Keys}

Root signing keys anchor the trust domain. They should be protected with the
strongest operational controls in the deployment. Rotation must be planned so
that relying parties can distinguish old valid Root records from forged
records. A production design should publish key identifiers, validity windows,
and rotation records.

\subsection{Registrar and Discovery Keys}

Registrar keys sign upstream submissions and credentials. Discovery keys sign
query responses. Both roles should support revocation and rotation. Root
authorization state should refer to current infrastructure DIDs and verification
methods, so that deactivated keys stop being trusted.

\subsection{Resource and Controller Keys}

Resource keys prove subject control and, for callable resources, sign runtime
requests and responses. Controller keys authorize the applicable registration
and identity-update operations. An external controller relationship does not
make the controller the signer of the resource's business messages.
Local OAN Identity material may contain private keys, but public DID
Documents, VCs, ResourcePackages, Discovery responses, and identity-directory
entries must not expose private key material. Key rotation should be handled as
an identity update. Root should accept the new identity version only after
verifying the complete updated representation and supporting evidence.

\section{Canonicalization and Hashing}

OAN relies on hash binding across protocol objects. If two
implementations serialize the same logical object differently, signatures and
hashes may fail. Protocol objects should therefore define canonical JSON
serialization or another deterministic representation. The following objects
especially require stable canonicalization:

\begin{itemize}[leftmargin=*]
  \item DID-style identity representation;
  \item DID Document proof input and final DID Document hash;
  \item registration credential unsigned payload;
  \item signed upstream request envelope;
  \item Root-verified package metadata;
  \item Discovery response payload;
  \item trusted invocation envelope.
\end{itemize}

Test vectors should include canonical byte strings and expected hashes. This
will reduce the risk that independent SDKs become incompatible despite using
the same field names.

\section{SDK Architecture}

SDKs encapsulate the security-sensitive operations required by OAN. They
should expose high-level functions while allowing applications and operators
to inspect verification results and failure reasons.

\subsection{Resource and Agent SDK}

The Resource or Agent SDK should help resource providers and consumers create
or import OAN Identity material, generate DID Documents, create signed
envelopes, verify peer DID Documents, verify credentials, check timestamps and
nonces, verify Discovery responses, and produce signed responses. It should
not hide verification failures behind vague exceptions, and it must keep local
private-key material separate from public DID Documents and credentials.

\subsection{Discovery SDK}

The Discovery SDK should help clients query Discovery nodes and verify signed
responses. It should return both candidate data and verification status. A
developer should be able to distinguish ``no candidates matched'' from
``Discovery response failed verification''.

Client configuration such as a \texttt{baseUrl} is a routing convenience, not a
trust authority. An SDK may use it to choose a Registrar, Discovery node, or
public portal endpoint, but it must still verify DID documents, Root package
evidence, Discovery signatures, timestamps, versions, and authorization-derived
status. Changing an endpoint should not silently change the trust decision
unless the returned protocol evidence also changes.

\subsection{Infrastructure SDK}

The Infrastructure SDK should support Root, Registrar, Discovery, and CDN
administration. It should build signed upstream requests, verify Root packages,
interact with authorization-domain vocabularies, and support conformance tests.
Governance operation support for the on-chain governance layer should be
isolated in designated governance clients or processes.
Default Registrar, Discovery, and VC issuer runtimes should depend on event
subscription and authorization-cache APIs rather than proposal or voting APIs.

\section{Deployment Topologies}

\subsection{Local Development}

Local development can run one Root, one Registrar, one Discovery node, one CDN,
one Service Agent, and one User Agent on a single machine. This topology
supports reproduction of lifecycle tests, negative cases, and small benchmarks.

\subsection{Single-Domain Pilot}

A single-domain pilot may separate Root, Registrar, Discovery, and CDN into
different services or machines while keeping one operator. This topology tests
network behavior, synchronization, package distribution, and operational
observability.

\subsection{Reference Public Deployment Profile}

The reference public deployment profile is a non-normative technical profile
that demonstrates how the reference roles can be exposed in a publicly
observable trust domain. A typical profile contains public interaction
surfaces, Root, Registrar, Discovery, package distribution, governance-state
indexing, persistent storage, and optional semantic-indexing components. The
profile may be deployed in many physical forms and should not be interpreted as
the only deployment shape.

The profile also demonstrates multiple Registrar and Discovery nodes in one
trust domain. These nodes follow the same governance, authorization-domain,
registration, distribution, synchronization, and Discovery rules. A node used
for public pilot validation is not a distinct protocol role. It is an instance
of the same Registrar or Discovery role with its own authorization and runtime
state.

Deployment-specific hostnames, cloud-provider choices, internal ports,
filesystem paths, account names, and secrets are outside the technical
specification and should remain in operator-controlled runbooks.

\subsection{Multi-Operator Trial}

A multi-operator trial combines independently operated Registrars and
Discovery nodes under shared verification rules. Root authorization and
authorized-domain boundaries preserve consistent admission and visibility
decisions while allowing each operator to manage its own infrastructure.

\subsection{Production Trust Domain}

A production trust domain should add durable databases, key management systems,
rate limits, monitoring, backup, disaster recovery, release signing, and
operator procedures. It should also define who can change Root authorization
state and how emergency revocation is performed.

\section{Observability}

Observability should be designed around lifecycle and trust events rather than
only HTTP status codes. Operators should track:

\begin{itemize}[leftmargin=*]
  \item Registrar draft state transitions;
  \item subject-control challenge success and failure;
  \item credential issuance counts and failures;
  \item Root acceptance and rejection reasons;
  \item Root publication queue depth and latency;
  \item CDN package availability and manifest history;
  \item Discovery synchronization cursor lag;
  \item Discovery package rejection reasons;
  \item query latency, result counts, and pagination behavior;
  \item nonce replay rejection and timestamp failures.
\end{itemize}

These metrics help distinguish performance bottlenecks from trust failures.
For example, a missing Discovery result may be caused by query mismatch,
publication lag, out-of-domain authorization, failed package verification, or
Root rejection. Observability should make these causes separable.

\subsection{Public Metrics Reporting}

Public metrics reporting is a visibility profile, not a trust source. Registrar
and Discovery nodes may publish or report public counters to a transparency
surface. That surface may aggregate the latest node snapshots for human
inspection, but clients must not treat those aggregates as authorization or
acceptance facts.

Registrar metrics may include node identity, role, region, total registered
resources, resources by type, and new registrations by time window. Discovery
metrics may include node identity, role, region, visible resources, Discovery
calls by time window, structured index state, semantic index state, and sync
freshness. These fields support transparency, but they do not decide
whether a resource is accepted, current, or callable.

\begin{figure*}[!ht]
\centering
\includegraphics[width=0.87\textwidth]{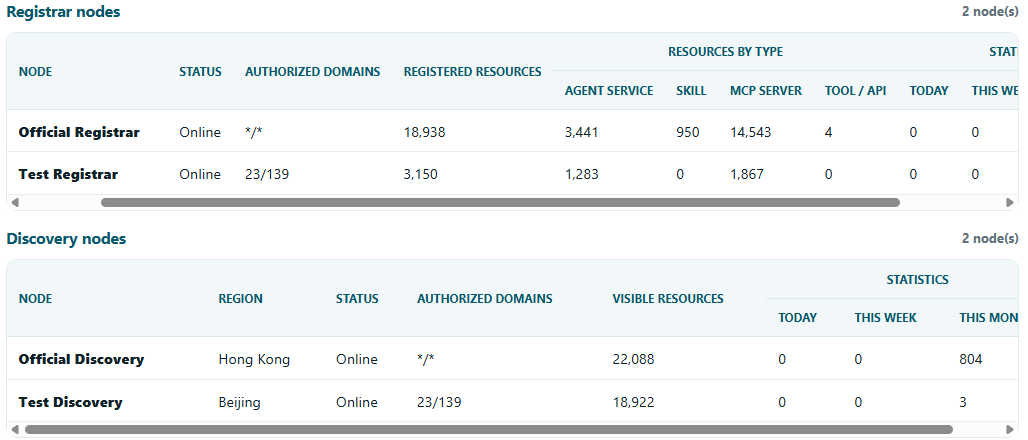}
\caption{Public node-level metrics as a non-normative visibility surface. The
underlying protocol truth remains in Root acceptance, package verification, and
Discovery authorization.}
\label{fig:yp-node-metrics}
\end{figure*}

Figure~\ref{fig:yp-node-metrics} illustrates this operational visibility;
the displayed counters supplement inspection without replacing protocol evidence.

\subsection{Public Governance Event View}

The public governance-event view is a curated transparency surface. It may show
event time, affected node, and human-readable event description. It should not
be used as runtime authorization input. Runtime authorization should be derived
from indexed governance state, Root-issued infrastructure authorization
material, and the verification logic of Registrar, Discovery, and relying
clients. A human-readable event table is therefore explanatory, not normative.

\section{Formal Invariants}

The following invariants summarize the technical design.

\subsection{I1: Root Acceptance Integrity}

If Root accepts an identity version, then the accepted record must bind the
resource DID, version, representation hash, status, and Root proof. Any later
package claiming that acceptance must match this bound hash.

\subsection{I2: Registrar Authorization}

If Root accepts a Registrar-submitted identity, then the submitting Registrar
must have been authorized at submission time and the signed upstream envelope
must verify under the Registrar identity.

\subsection{I3: Discovery Scope}

If a compliant Discovery node indexes or exposes a resource identity, then
the node and resource must satisfy the \(DiscoveryScope(d,M)\) predicate
defined in the Discovery Authorization section. Capability tags may refine
matching and ranking, but they must not expand Discovery authorization.

\subsection{I4: Current-Version Visibility}

If a resource identity version has been superseded or revoked and Discovery has
processed the relevant Root state, then that version must not remain visible as
the current admissible version.

\subsection{I5: Invocation Binding}

If a Service Agent accepts a trusted invocation envelope, then the envelope
signature, timestamp, nonce, target DID, body hash, and caller credential
binding must have passed verification.

\subsection{I6: DID Proof and Controller Authorization}

If a DID Document is accepted into the OAN trust path, then its DID Document
proof must verify under the declared verification method. If the resource is
externally controlled, the controller authorization proof must also bind the
resource, controller, Registrar, signed document hashes, purpose, validity
window, and nonce. For an existing resource, the authorizing controller is the
one recorded in the current Root-accepted version. A proposed successor cannot
authorize its own appointment, and an update from a superseded controller must
be rejected.

\subsection{I7: Root Identity Directory Non-Authority}

If an application uses Root's identity directory for login public-key
discovery, then a directory hit is not sufficient by itself. The application
must still verify a fresh challenge signature under the listed public key. The
directory does not authorize resource control or replace DID Document and VC
verification.

If the same DID is used as a controller for another resource, that registration
still requires the controller authorization proof defined for resource
registration. Login success and controller authorization are intentionally
different protocol decisions.

\subsection{I8: Semantic Recovery}

If structured synchronization succeeds but semantic indexing fails, the failure
must leave durable retry state. A compliant Discovery node must not silently
drop the semantic-indexing work item after accepting the verified package into
its structured state.

\subsection{I9: Public Surface Non-Authority}

If a public portal, public metrics cache, or curated governance-event table is
stale or unavailable, protocol truth must still be derived from Root
acceptance, indexed governance state, package verification, Discovery
authorization, and signed response verification.

\section{Backward and Forward Compatibility}

OAN should evolve without breaking every implementation. Compatibility can be
managed through typed objects, optional fields, protocol version fields,
capability negotiation, and conformance profiles. Critical verification fields
should be mandatory. Experimental metadata should be optional and must not
change trust semantics unless a new profile explicitly says so.

A useful compatibility rule is: old verifiers may ignore unknown non-critical
metadata, but they must reject objects that require unknown critical trust
semantics. This prevents silent downgrade when new security-relevant fields are
introduced.

This rule is especially important for schema and directory integration. An
external schema record, ADS-style directory hint, marketplace descriptor, or
runtime-specific extension can be carried as non-critical metadata when it only
improves description, search, or routing. It must become a critical extension
only when a verifier is expected to use it for trust decisions, such as
authorization scope, endpoint authenticity, version validity, revocation, or
credential binding. This keeps OAN extensible without letting ecosystem
metadata accidentally redefine the trust model.

\subsection{Wire-Format Stability}

Wire-format specifications and versioned conformance vectors give concrete
form to the semantic contracts described here. Interoperability requires
agreement on canonical JSON rules, required and optional fields, field naming,
object version identifiers, signature preimages, hash algorithms, HTTP methods
and paths, pagination tokens, timestamp formats, nonce scope, error codes, and
critical-extension markers.

Independent implementations use the applicable object profiles and conformance
vectors to establish compatibility. A breaking field change requires a new
object version or conformance profile. A non-breaking extension remains
optional; verifiers ignore unrecognized non-critical extensions. An extension
that changes a trust decision must be explicitly marked as critical and
supported by the verifier.

\section{Reference Verification Algorithms}

The following algorithms describe expected verifier behavior at a high level.
They are not tied to one programming language.

\subsection{Root Submission Verification}

Root submission verification should proceed in a fail-closed order. First,
parse the request and reject malformed objects. Second, verify the upstream
envelope signature and body hash. Third, check timestamp freshness and nonce
uniqueness. Fourth, resolve the Registrar identity and confirm active
authorization. Fifth, verify the registration credential, including issuer,
subject, validity, and proof. Sixth, verify DID document structure,
verification methods, capability metadata, and subject binding. Seventh, check
version transition rules and, for an existing resource, verify the update under
the controller recorded in the current accepted state. Only after these checks
should Root atomically create the accepted record and advance the corresponding
controller state.

\subsection{Discovery Package Verification}

Discovery package verification should parse the package, recompute hashes,
verify Root proof, check bulletin evidence, check current-version state, and
apply authorized-domain authorization. Only packages that pass all checks should
enter the query index. Failed packages should be retained in a rejected-package
store with reason codes so that operators can debug synchronization and policy
issues.

\subsection{Discovery Backfill Verification}

Discovery backfill should proceed as a repeated verification loop rather than
as a blind database copy. First, load the Discovery node's current authorization
state and synchronization cursor. Second, fetch candidate package identifiers
or cursor ranges from Root or the distribution layer. Third, verify each
package with the same package-verification algorithm used for incremental
synchronization. Fourth, apply authorized-domain filtering before writing
structured index state. Fifth, enqueue semantic indexing for accepted
structured records. Sixth, record semantic success, pending, or failure state
durably. Seventh, advance the cursor only according to the implementation's
chosen cursor safety rule. Repeating these steps should be idempotent.

Backfill must not bypass the normal trust path. It should not import resources
that fail Root proof, current-version checks, lifecycle checks, or Discovery
authorization. It should also not treat semantic-index completion as a
condition for Root acceptance. Semantic indexing affects intent matching and
ranking; Root acceptance and Discovery authorization remain the trust gates.

\subsection{User Agent Candidate Verification}

A User Agent should verify the Discovery response signature, timestamp,
Discovery authorization state, selected candidate provenance, Root package
hashes, and current-version status before constructing an invocation. The User
Agent should not treat Discovery output as plain search output. It is signed
trust evidence that still requires validation.

\subsection{Service Agent Invocation Verification}

A Service Agent should verify caller identity, caller credential, target DID,
request path, body hash, timestamp, nonce, and request proof. It should also
apply local authorization policy after OAN checks pass. OAN proves who the
caller claims to be and whether the request is fresh and bound; local policy
decides whether that caller may perform the requested business action.

\section{Pagination, Batching, and Backpressure}

The reference experiments show that large-result queries, publication, and
synchronization are important bottleneck areas. The architecture should
therefore treat pagination, batching, and backpressure as protocol-relevant
engineering concerns.

\subsection{Pagination}

Discovery query responses should support pagination when result sets become
large. Pagination metadata must be covered by the Discovery response signature
so that a client can verify page boundaries and query context. A page token
should not allow a client to bypass authorization checks or retrieve stale
results outside the original query policy.

\subsection{Publication Batching}

Root-to-CDN and Root-to-Discovery publication can be batched. Batch records
should include cursor ranges, package identifiers, status, retry count, and
error state. Batching should improve throughput without hiding individual
package verification failures.

\subsection{Backpressure}

Registrar submission paths should have explicit backpressure. A burst of draft
submissions should not create unbounded Root requests or local contention.
Backpressure can include bounded queues, single-flight protection per draft,
timeouts, retry policies, and clear submission-state reporting.

\section{Degraded Behavior}

Production systems must define behavior during partial failure.

\subsection{Root Unavailable}

If Root is unavailable, Registrar may continue draft preparation and local DID
control checks, but it cannot claim Root acceptance. Discovery should continue
serving previously verified current data within freshness policy, but it should
not invent new Root facts.

\subsection{CDN Unavailable}

If CDN is unavailable, Discovery synchronization may lag. Existing verified
index data may remain usable within policy, but new packages will not become
visible until synchronization resumes. Operators should expose cursor lag and
sync failures.

\subsection{Discovery Stale}

If Discovery is stale, User Agents may see incomplete results. This is safer
than accepting unverifiable data. Discovery should publish status that allows
clients and operators to understand synchronization lag.

\subsection{Nonce Store Failure}

If a Service Agent cannot check nonce uniqueness, it should fail closed for
high-risk operations. For low-risk reference scenarios, a degraded mode may exist,
but it should be explicit and should not be presented as full trusted
invocation.

\section{Data Retention}

OAN deployments should define retention policies for drafts, credentials,
accepted records, rejected submissions, package history, Discovery rejection
logs, query logs, and invocation evidence. Retention should balance audit
requirements with privacy and operational cost.

Root accepted records and revocation history usually require longer retention
because they define trust-domain history. Registrar drafts and private review
evidence may need shorter or policy-specific retention. Discovery query logs
may contain sensitive interest patterns and should be minimized or protected.
Invocation logs should be controlled by the Service Agent's business domain and
privacy obligations.

\section{Release and Supply-Chain Integrity}

The OAN software supply chain should be treated as part of the trust model.
Signed releases, reproducible build metadata, dependency review, artifact
hashes, and compatibility statements help operators know which implementation
they are running. Release practices should distinguish protocol-compatible
changes from experimental or breaking changes.

Reference services should publish version information and protocol object
support. SDKs should publish test-vector compatibility. Trial networks should
record which service versions are active so that experiments and reference scenarios can be
reproduced.

\section{Operational Security Baseline}

Production deployment requires more than passing reference integration tests.
An OAN trust domain should define at least the following controls:

\begin{itemize}[leftmargin=*]
  \item durable nonce stores and timestamp skew policy for privileged writes
  and trusted invocation;
  \item event synchronization lag monitoring for Registrar, Discovery, and VC
  issuer authorization state;
  \item fail-closed authorization cache behavior when bulletin or Root state is
  unavailable or too stale;
  \item package verification failure alerts and rejected-package retention;
  \item Discovery index rebuild, cursor recovery, and query-result pagination;
  \item bounded queues, backpressure, retry policy, and timeout policy for
  Registrar submission and Root publication paths;
  \item key rotation, key compromise response, and operator access-control
  procedures;
  \item audit log retention for Root decisions, Registrar submissions,
  Discovery synchronization, and trusted invocation failures;
  \item release verification, dependency review, and conformance-test records
  for deployed service and SDK versions.
\end{itemize}

These controls do not change the protocol model, but they determine whether an
OAN deployment can be trusted operationally by other organizations.

\section{Test Vectors and Profiles}

Interoperability requires more than prose. OAN implementations should publish
test vectors that allow another implementation to verify whether it computes
the same hashes, signatures, and authorization decisions.

\subsection{Core Test Vectors}

Core test vectors should include a canonical DID-style identity representation,
its canonical byte sequence, its expected hash, a valid registration credential,
an invalid credential with a wrong subject, a valid upstream envelope, an
envelope with a mismatched body hash, a Root-verified package, a Discovery
response, and a trusted invocation envelope. Each vector should include the
expected verification result and the reason for rejection when applicable.

Additional vectors should cover resource and recovery semantics. These include
a resource type mismatch between
\texttt{did:oan} subject code and product metadata, an unauthorized Registrar
submission, a Discovery node attempting to expose an out-of-domain resource, a
revoked or superseded package, a Root proof mismatch, a semantic-index failure
that leaves pending retry state, a repeated backfill operation that remains
idempotent, and a stale public metrics cache that does not change protocol
verification results.

\subsection{Conformance Profiles}

OAN can define conformance profiles for different implementation depths. A
basic identity profile may only parse and verify identity documents and Root
packages. A Discovery profile must additionally enforce authorized-domain
authorization and sign responses. A trusted invocation profile must verify
caller credentials, nonces, timestamps, target binding, body hashes, and
request signatures. A full infrastructure profile must implement Root,
Registrar, Discovery, and CDN semantics.

An operator profile should additionally preserve durable state, expose
operational freshness, support Discovery backfill, and retain recoverable
semantic-index failures. Public metrics may be included as a visibility
extension, but they are not a prerequisite for the wire-format identity profile.
These requirements test operational compatibility rather than wire-format
identity alone.

\subsection{Security Configuration Baseline}

A production profile should define minimum security configuration. Examples
include nonce retention windows, timestamp skew tolerance, maximum request body
hash age, key rotation procedure, rejected-submission logging, Discovery sync
lag alerts, package verification failure alerts, and disabled degraded modes
for high-risk operations. These values may vary by deployment, but the presence
of explicit configuration is itself part of operational maturity.

\subsection{Interoperability Reports}

Independent implementations can publish interoperability reports to make
compatibility visible and actionable.
Such reports should state which profiles are supported, which test vectors
passed, and which optional extensions are implemented. Shared reports and test
vectors provide a practical basis for interoperable implementations and
progressive profile adoption.

\section{Privacy and Data Minimization}

OAN identity representations are intended to be discoverable, but that does not
mean every piece of operator evidence should be public. Private review evidence
should remain in governance workflow or Registrar records unless there is a
clear reason to publish it. Public packages should contain the material needed
for verification and discovery, not unnecessary personal or operational data.

The separation among Root acceptance, package publication, and Discovery
exposure provides a foundation for privacy extensions such as confidential
queries, selective disclosure, private capability matching, scoped credentials,
and policies for sensitive Agent categories.

\section{Open Technical Work}

The reference implementation establishes the core identity lifecycle. The
remaining technical agenda covers protocol breadth, deployment, and ecosystem
trust services:

\begin{itemize}[leftmargin=*]
  \item MCP and A2A conformance for asynchronous tasks, streaming, callbacks,
  cancellation, and resource or artifact lifecycles;
  \item production key custody, rate limiting, recovery drills, and operational
  security baselines;
  \item independent implementation test vectors and interoperability reports;
  \item richer semantic Discovery, taxonomy governance, backfill, pagination,
  and result inspectability;
  \item interoperable interfaces for third-party credential issuers and trust
  bridges, including issuer authorization, status, revocation, validity, and
  evidence composition;
  \item privacy-preserving Discovery and confidential capability queries;
  \item signed release tooling, reproducible artifact verification, and
  multi-root or cross-trust-domain federation.
\end{itemize}

\section{Conclusion}

OAN defines a verifiable lifecycle for governed resource identity and trusted
Discovery, combining Root-enforced provenance, current-version state,
authorization-scoped exposure, package verification, and signed
pre-connection validation. Heterogeneous protocols can use this lifecycle
without adopting one task protocol or runtime. Independent implementations
remain interoperable when they make the same trust-boundary decisions and pass
common test vectors; deployment and storage choices may differ. The L0--L3
model supports incremental adoption without changing the underlying resource
identity semantics.

The current evidence establishes structured lifecycle completion for 60
isolated runs up to approximately 5000 resources and exercises the security
properties mapped to I1--I7. Semantic-index convergence, physical distributed
scaling, multi-Root consistency, and live interoperability for the three
application scenarios remain unverified.

\section*{Acknowledgment}

The author thanks Jin Jian, Xie Jiagui, Li Bingqi, and Zhu Runkai for their
support.

\nocite{xu2026grail,xu2026darwinnet,wu2023autogen,wang2024llmagentsurvey,terry1995managing,rfc5280,myers1999ocsp,ruan2024toolemu}
\bibliographystyle{IEEEtran}
\bibliography{Ref}

\begin{thebibliography}{10}
\providecommand{\url}[1]{#1}
\csname url@samestyle\endcsname
\providecommand{\newblock}{\relax}
\providecommand{\bibinfo}[2]{#2}
\providecommand{\BIBentrySTDinterwordspacing}{\spaceskip=0pt\relax}
\providecommand{\BIBentryALTinterwordstretchfactor}{4}
\providecommand{\BIBentryALTinterwordspacing}{\spaceskip=\fontdimen2\font plus
\BIBentryALTinterwordstretchfactor\fontdimen3\font minus
  \fontdimen4\font\relax}
\providecommand{\BIBforeignlanguage}[2]{{%
\expandafter\ifx\csname l@#1\endcsname\relax
\typeout{** WARNING: IEEEtran.bst: No hyphenation pattern has been}%
\typeout{** loaded for the language `#1'. Using the pattern for}%
\typeout{** the default language instead.}%
\else
\language=\csname l@#1\endcsname
\fi
#2}}
\providecommand{\BIBdecl}{\relax}
\BIBdecl

\bibitem{w3c_did_core}
{W3C Credentials Community Group}, ``Decentralized identifiers (dids) v1.0,''
  \url{https://www.w3.org/TR/did-core/}, 2022.

\bibitem{vc_data_model_2}
{W3C Verifiable Credentials Working Group}, ``Verifiable credentials data model
  v2.0,'' \url{https://www.w3.org/TR/vc-data-model-2.0/}, 2025.

\bibitem{vc_data_integrity}
------, ``Verifiable credential data integrity 1.0,''
  \url{https://www.w3.org/TR/vc-data-integrity/}, 2024.

\bibitem{did_resolution}
{W3C Credentials Community Group}, ``Did resolution,''
  \url{https://w3c-ccg.github.io/did-resolution/}, 2024.

\bibitem{mcp_spec}
{Model Context Protocol Contributors}, ``Model context protocol
  specification,'' \url{https://modelcontextprotocol.io/specification/}, 2025.

\bibitem{a2a_spec}
{A2A Protocol Contributors}, ``Agent2agent protocol specification,''
  \url{https://a2a-protocol.org/latest/specification/}, 2025.

\bibitem{did_wba_spec}
{Agent Network Protocol Contributors}, ``Did:wba method specification:
  Web-based agent decentralized identifier,''
  \url{https://agentnetworkprotocol.com/en/specs/03-did-wba-method-specification/},
  2025.

\bibitem{rose2020zerotrust}
S.~Rose, O.~Borchert, S.~Mitchell, and S.~Connelly, ``Zero trust
  architecture,'' NIST Special Publication 800-207, 2020.

\bibitem{spiffe_spire}
{Cloud Native Computing Foundation}, ``The spiffe and spire project for
  workload identity,'' \url{https://spiffe.io/}, 2022.

\bibitem{gruner2023ssi}
A.~Gr{\"u}ner, A.~M{\"u}hle, and C.~Meinel, ``Analyzing and comparing the
  security of self-sovereign identity management systems through threat
  modeling,'' in \emph{Proceedings of the 38th ACM/SIGAPP Symposium on Applied
  Computing}, 2023, pp. 389--396.

\bibitem{glockler2023iam}
M.~Gl{\"o}ckler, J.~Sedlmeir, V.~Schlatt, and N.~Urbach, ``A systematic review
  of identity and access management requirements in enterprises and potential
  contributions of self-sovereign identity,'' \emph{Business \& Information
  Systems Engineering}, vol.~65, no.~4, pp. 421--440, 2023.

\bibitem{popa2023chaindiscipline}
M.~Popa, X.~Yue, A.~Boudguiga, A.~Chibani, and H.~B{\'e}cha, ``Chaindiscipline:
  A blockchain-based self-sovereign identity framework for iot,'' \emph{IEEE
  Transactions on Services Computing}, vol.~16, no.~4, pp. 2571--2584, 2023.

\bibitem{agntcy_oasf}
{AGNTCY Contributors}, ``{Open Agentic Schema Framework},''
  \url{https://github.com/agntcy/oasf}, 2026, accessed: 2026-06-05.

\bibitem{xu2026grail}
J.~Xu, ``Grail: A deep-granularity hybrid resonance framework for real-time
  agent discovery via slm-enhanced indexing,'' \emph{arXiv preprint
  arXiv:2605.02489}, 2026.

\bibitem{mcp_registry}
{Model Context Protocol Contributors}, ``{Model Context Protocol Registry},''
  \url{https://github.com/modelcontextprotocol/registry}, 2026, accessed:
  2026-06-05.

\bibitem{agntcy_ads_overview}
{AGNTCY Contributors}, ``{Agent Directory Service Overview},''
  \url{https://docs.agntcy.org/dir/overview/}, 2026, accessed: 2026-06-05.

\bibitem{huang2025ans}
K.~Huang, V.~S. Narajala, I.~Habler, and A.~Sheriff, ``Agent name service
  (ans): A universal directory for secure ai agent discovery and
  interoperability,'' \emph{arXiv preprint arXiv:2505.10609}, 2025.

\bibitem{xu2026darwinnet}
J.~Xu and B.~Li, ``Darwinnet: An evolutionary network architecture for
  agent-driven protocol synthesis,'' \emph{arXiv preprint arXiv:2604.01236},
  2026.

\bibitem{wu2023autogen}
Q.~Wu, G.~Bansal, J.~Zhang, Y.~Wu, B.~Li, E.~Zhu, L.~Jiang, X.~Zhang, S.~Zhang,
  J.~Liu, A.~H. Awadallah, R.~W. White, D.~Burger, and C.~Wang, ``Autogen:
  Enabling next-gen llm applications via multi-agent conversation,''
  \emph{arXiv preprint arXiv:2308.08155}, 2023.

\bibitem{wang2024llmagentsurvey}
L.~Wang, C.~Ma, X.~Feng, Z.~Zhang, H.~Yang, J.~Zhang, Z.~Chen, J.~Tang,
  X.~Chen, Y.~Lin, W.~X. Zhao, Z.~Wei, and J.-R. Wen, ``A survey on large
  language model based autonomous agents,'' \emph{Frontiers of Computer
  Science}, vol.~18, no.~6, p. 186345, 2024.

\bibitem{terry1995managing}
D.~B. Terry, M.~M. Theimer, K.~Petersen, A.~J. Demers, M.~J. Spreitzer, and
  C.~H. Hauser, ``Managing update conflicts in bayou, a weakly connected
  replicated storage system,'' \emph{Proceedings of SOSP}, pp. 172--183, 1995.

\bibitem{rfc5280}
D.~Cooper, S.~Santesson, S.~Farrell, S.~Boeyen, R.~Housley, and T.~Polk,
  ``Internet x.509 public key infrastructure certificate and certificate
  revocation list (crl) profile,'' RFC 5280, 2008.

\bibitem{myers1999ocsp}
M.~Myers, R.~Ankney, A.~Malpani, S.~Galperin, and C.~Adams, ``X.509 internet
  public key infrastructure online certificate status protocol - ocsp,'' RFC
  2560, 1999.

\bibitem{ruan2024toolemu}
\BIBentryALTinterwordspacing
Y.~Ruan, H.~Dong, A.~Wang, S.~Pitis, Y.~Zhou, J.~Ba, Y.~Dubois, C.~J. Maddison,
  and T.~Hashimoto, ``{ToolEmu}: Identifying the risks of lm agents with an
  lm-emulated sandbox,'' in \emph{The Twelfth International Conference on
  Learning Representations}, 2024. [Online]. Available:
  \url{https://openreview.net/forum?id=NB5FP1A2yC}
\BIBentrySTDinterwordspacing

\end{thebibliography}

\end{document}